\documentclass[iop,apj]{emulateapj}
\usepackage{amsmath}
\usepackage{adjustbox}

\usepackage{apjfonts}
\usepackage{graphicx}
\usepackage{ifpdf}
\usepackage{natbib}

\newcommand\lambdap{\lambda_{\rm p}}
\newcommand\omegap{\omega_{\rm p}}

\newcommand\degrees{\,^{\circ}}

\begin{document}
\title{Beaming of particles and synchrotron radiation in relativistic magnetic reconnection}

\author{Daniel Kagan\altaffilmark{1,2}, Ehud Nakar\altaffilmark{1}, Tsvi Piran \altaffilmark{2}}
\email{daniel.kagan@mail.huji.ac.il}
\affil{$^1$Raymond and Beverly Sackler School of Physics \& Astronomy,  Tel Aviv University, Tel Aviv 69978, Israel\\
$^2$Racah Institute of Physics, The Hebrew University, Jerusalem 91904, Israel}

\begin{abstract}Relativistic reconnection has been invoked as a mechanism for particle acceleration in numerous astrophysical systems.  According to idealised analytical models reconnection produces a bulk relativistic outflow emerging from the reconnection sites (X-points). The resulting radiation is therefore highly beamed. Using two-dimensional particle-in-cell (PIC) simulations, we investigate particle and radiation beaming, finding a very different picture. Instead of having a relativistic average bulk motion with isotropic electron velocity distribution in its rest frame, we find that the bulk motion of particles in X-points is similar to their Lorentz factor $\gamma$, and the particles are beamed within $\sim5/\gamma$.   On the way from the X-point to the magnetic islands, particles turn in the magnetic field,  forming a fan confined to the current sheet.  Once they reach the islands they isotropise after completing a full Larmor gyration and their radiation is not strongly beamed anymore. The radiation pattern at a given frequency depends on where the corresponding emitting electrons radiate their energy.  Lower energy particles that cool slowly spend most of their time in the islands, and their radiation is not highly beamed. Only particles that quickly cool at the edge of the X-points generate a highly beamed fan-like radiation pattern.  The radiation emerging from these fast cooling particles is above the burn-off limit ($\sim100 $ MeV in the overall rest frame of the reconnecting plasma.) This has significant implications for models of GRBs and AGNs that invoke beaming in that frame at much lower energies.  \end{abstract}

\section{Introduction}\label{sec:introduction} Magnetic reconnection is thought to be one of the most important mechanisms by which magnetic energy is converted to kinetic energy in astrophysical plasmas \citep[for a review, see e.g.][]{yamada_10}. This conversion may produce a tail in the particle energy distribution including randomly oriented relativistic particles and high-frequency radiation,  coherent bulk outflows, or both. In the relativistic reconnection regime, in which the ratio of the magnetic energy to the total enthalpy of the particles (the magnetization $\sigma$) is much larger than 1, magnetic reconnection has been hypothesised to be responsible for observed high-energy emission in the prompt phase of gamma-ray bursts (GRBs) \citep{thompson_94, lyutikov_03,giannios_05,lyutikov_06,icmart_11, mckinney_12}, emission from pulsar wind nebulae \citep{kirk_03,  sironi_11,petri_12}, and active galactic nucleus (AGN) jets \citep{2009MNRAS.395L..29G,giannios_13,nalewajko_11,narayan_12}. Relativistic jets launched by the Blandford-Znajek mechanism \citep{blandford_zn}  are expected to be dominated by magnetic energy, so reconnection is a primary mechanism invoked for the conversion of these fields to particles and radiation. Since relativistic bulk outflows have been predicted to occur in relativistic reconnection \citep{2005MNRAS.358..113L}, it has been hypothesized that the variability of these sources may arise from the direction of these outflows swinging past the observer's line of sight. The fast variability model of blazar jets \citep{2009MNRAS.395L..29G,giannios_13,nalewajko_11,narayan_12}, and the jet-within-a-jet model for the prompt emission in GRBs (assuming that the jet is magnetically dominated) \citep{narayan_09,lazar_09,zhang_14, binyamini_15} both rely on this beaming to produce the fast variability observed, so the amount of beaming produced in magnetic reconnection is critical for evaluating their viability.

Numerical investigations of particle acceleration in relativistic magnetic reconnection are generally carried out using particle-in-cell (PIC) simulations that include all kinetic effects required to probe the non thermal spectrum of high-energy particles and the possible anisotropy in the particle distribution. These simulations, which have mostly focused on the pair-plasma case which is easier to simulate, have shown that reconnection can produce non thermal tails of highly accelerated particles in both two and three dimensions \citep{{2001ApJ...562L..63Z,zenitani_05b,zenitani_07, zenitani_hesse_08b,jaroschek_04,jaroschek_08b,bessho_05,bessho_07,bessho_12,daughton_07,lyubarsky_liverts_08, liu_11, cerutti_12b, cerutti_13a, cerutti_14, werner_16, 2014ApJ...783L..21S,sironi_16,guo_14,guo_15,liu_15}}, and results are similar when electron-ion plasmas are studied \citep{melzani_14, sironi_15, guo_16}. In most cases, the spectrum of the highly accelerated particles has been consistent with a power law, although analytical models by \citet{larrabee_03} and \citet{bessho_12} indicate that the distribution may be different.  This distribution can be hard enough, especially at large magnetisation, that the particles at the highest Lorentz factors dominate the energy distribution \citep{2014ApJ...783L..21S, melzani_14, werner_16,guo_14,guo_15,guo_16}. It has been suggested that the drift-kink instability (DKI) may pnt the development of fast reconnection and particle acceleration in three-dimensional simulations without a guide field perpendicular to the reversing field of reconnection \citep{zenitani_08}, but more recent work has shown that on longer timescales fast reconnection and particle acceleration indeed take place, just as in the two-dimensional case \citep{2014ApJ...783L..21S}. In this work, we therefore carry out two dimensional simulations with the expectation that the results will apply to three-dimensional reconnection as well. We do note that the details of reconnection in 2D are likely to be different from those of the reconnection on long timescales that follows the initial effects of the drift-kink instability, but the overall physics is likely to be the same. 
The relativistic beaming of particles and the resulting synchrotron radiation resulting from coherent bulk outflows in reconnection has been investigated to a lesser degree.  The evidence for the presence of relativistic bulk flows in simulations is uncertain, with \citet{2014ApJ...783L..21S} finding fast relativistic flows, and  \citet{melzani_14} finding in the case of electron-ion plasmas that only mildly relativistic outflows are present. { In relativistic MHD simulations of forced reconnection with moderate magnetization, \citet{deng_15} found that near-Alfv{\'e}nic outflows were produced, although it is uncertain whether this would continue at high magnetization}.  It should be noted that the bulk outflows may not be a good probe of the beaming of the outgoing radiation, since the hard power laws found in relativistic reconnection indicate that the output flux will be dominated by higher-energy particles, which may not be beamed in the same way as the bulk plasma. \citet{cerutti_12b, cerutti_13a,cerutti_14} have found in both two and three-dimensional simulations that substantial beaming of particles and radiation results from magnetic reconnection; the radiation can be beamed within $\sim 2-4\%$ of the sky,  However, these simulations focus on the radiation of particles close to the synchrotron burnoff limit, where they radiate in a time comparable to their period of gyration. \citet{yuan_16} carries out simulations of explosive reconnection including radiative feedback, finding significant beaming and variability in both the fast and slow cooling cases, but low radiative efficiency in the slow cooling case. In this work, we investigate the beaming of particles under the assumption that they do not radiate their energy quickly.

Relatively little numerical investigation has been done of relativistic reconnection at high magnetisations, but there is significant evidence that reconnection physics, and thus, beaming, will be significantly affected. \citet{bessho_12} and \citet{liu_15}, but not \citet{guo_14}, find that at high magnetisations the velocity of inflows into the current sheet and the normalised rate of reconnection of magnetic fields becomes extremely high, of order unity as opposed to the typical values of $0.05-0.2$ found for both nonrelativistic \citep{birn_01} and relativistic \citep[e. g. ][]{liu_11,bessho_12,melzani_14,liu_15} reconnection at lower magnetisation.  It is possible that at high magnetisations the density of background plasmas is very low. While magnetospheric space-plasma measurements have indicated that density contrasts in nonrelativistic reconnection are typically modest \citep{eastwood_average_2010}, relativistic reconnection has never been observed directly and it is uncertain whether this result will apply in the relativistic case. \citet{bessho_12} finds that in this low-density case, the current sheets became very thick, and there is a deficit of particles at intermediate Lorentz factors between the beginning of the nonthermal tail and the highest-energy particles. Our simulations investigate the physics of relativistic reconnection at high magnetisation and low background density, as well as extending the investigation of beaming to these relatively unexplored regimes.

This paper is organised as follows. In Section \ref{sec:methodology} we discuss our method of simulation and calculation of synchrotron radiation. In Section \ref{sec:results}, we present the results of our simulations, including our calculations of  the beaming of both particles and synchrotron radiation. Finally, in Section \ref{sec:conclusions}, we discuss our conclusions.
\section{Methodology} \label{sec:methodology}

We use the relativistic particle-in-cell (PIC) plasma code {\tt TRISTAN-MP} \citep{spitkovsky_structure_2008} to simulate the evolution of relativistic reconnection in two dimensions in a pair plasma.  Particle-in-cell simulations include the kinetic effects of individual particle motions by evolving the discretised versions of the exact equations of electrodynamics - Maxwell's Equations  and the Lorentz force law.  The fields are calculated on the vertices of a grid, while the particles that are evolved using the Lorentz force law (interpolated from the grid to their positions) are macroparticles, each of which represent many physical particles. This section describes the initial conditions, parameter choices, and resolution requirements in these simulations.

\subsection{Initial configuration}

We use a rectangular spatial domain of the form $0\leq x<L_x$, $0\leq y<L_y$, and the boundary conditions are periodic in all directions. The initial configuration contains two Harris current sheets \citep{harris62} without guide field at  $x=L_x/4$ and $x=3L_x/4$ with antiparallel currents. It is characterised by the magnetic field profile

\begin{equation}
  {\mathbf B}=B_0 \left[ \tanh \left(\frac{x-L_x/4}{\delta}\right) - \tanh \left(\frac{x-3L_x/4}{\delta}\right)-1\right]\hat{{\mathbf y}},
\label{eq:harris_field}
\end{equation}
where $\delta$ is the half-thickness of each current sheet. 

The density profile, which is defined including both species, consists of a specially varying, drifting current population entered at each current sheet, plus a background population of stationary particles.

\begin{equation}
\label{eq:density_profile}
n=n_0 \left[ {\rm sech}^2\left(\frac{x-L_x/4}{\delta}\right) +{\rm sech}^2\left(\frac{x-3L_x/4}{\delta}\right)\right]+n_{\rm b}.
\end{equation}

Pressure equilibrium between the background and the current sheet requires that $B_0^2=4\pi n_0 T_0$, where $T_0$ is the temperature of the particles (in units of $m c^2$) in the drifting current sheet population in the simulation frame. The density in the middle of each current sheet in the lab frame is $n_{\rm b}+n_0$, and the density in the background plasma is $n_{\rm b}$. 

 The drift velocity $\boldsymbol{\beta}$ of the positively charged and negatively charged particles in the current sheet population is given by the relation
\begin{equation}
\boldsymbol{\beta}_+=-\boldsymbol{\beta}_-=\pm B_0 /(4\pi n_0 q \delta) (-\hat{{\mathbf z}}),
\label{eq:drift}
\end{equation}
where the sign is positive for particles in the current sheet at $x=L_x/4$ and negative for particles in the current sheet at $x=3 L_x/4$.

We do not add any initial perturbation to this equilibrium, but instead allow instabilities to grow from noise. 
\subsection{Parameters}
One of the important parameters determining the physics of reconnection is the value of the magnetization  $\sigma$,  which in a pair plasma is given by
\begin{equation}
\sigma\equiv \frac{B^2}{4\pi n m c^2 h},
\label{eq:sigmadef}
\end{equation}
where $h=\gamma +P/(m n c^2)$ is the average enthalpy of particles,  $\gamma$ is the mean particle Lorentz factor and $P$ is the particle pressure.

We investigate the physics of reconnection at three values of the magnetisation in the background plasma $\sigma_{0}=$4, 40, and 400. 
We initialize the drifting and stationary populations of both species in a relativistic Maxwellian at temperature $T_0=0.5 m_e c^2$, which corresponds to $\gamma=2.05$ and $h=2.55$.  The value of $n_0/n_b$ used to initialise each simulation may be found using the equation, 

\begin{equation}
\frac{n_0}{n_b}=\frac{\sigma_0 h}{T_0},
\label{eq:nratio}
\end{equation}
where we define $\sigma_0$ as the magnetisation in the background plasma. This equation can be derived by combining the equation of pressure equilibrium above with the definition of  $\sigma$ (\ref{eq:sigmadef}). 

The other important parameter for the physics of reconnection is the ratio $\delta/\lambda_{\rm p}$, where $\lambda_{\rm p}$ is the plasma skin depth of particles in the center of the current sheet, given by 
\begin{equation}
\lambda_{\rm p}=\sqrt{\frac{\gamma m c^2}{4 \pi n_0 q^2 }}.
\label{eq:plasma}
\end{equation} 

This ratio sets the drift velocity of the plasma, which can be found by combining Equations (\ref{eq:drift}), (\ref{eq:nratio}), and  (\ref{eq:plasma}) to yield

\begin{equation}
\beta=\sqrt{\frac{T_0}{\gamma}} \frac{\lambda_{\rm p}}{\delta}.
\label{eq:othdrift}
\end{equation}

We choose to set $\delta/\lambda_{\rm p}=3$, which corresponds to a mildly relativistic drift velocity in the current sheet $\beta=0.17$. This current sheet width is small enough that fast reconnection can occur.

Our choices for the parameters set the physical system up to the choice of fiducial magnetic field $B_0$, so that it can be scaled to model many physical systems.

\subsection{Resolution requirements}
In a PIC simulation, the number of macroparticles of each species located in each grid cell must be large enough to resolve variations in the current density and limit high-frequency particle noise. {\tt TRISTAN-MP} uses a current filtering algorithm to reduce high-frequency particle noise, substantially reducing the required number of macroparticles per cell per species. A convergence test described in the work of \citet{kagan_13} has demonstrated that the evolution of reconnection configurations with this filtering in {\tt TRISTAN-MP} is unaffected by particle noise for densities as low as 4 particles/cell/species. We choose to initialize the simulations with 8 macroparticles/cell/species, but we have carried out convergence tests for particle densities up to $25 $ macroparticles/cell/species, finding similar results.  Density gradients in the initial configuration are accounted for without computational cost by varying the mass of macroparticles while keeping the charge-to-mass ratio the same. In all simulations, the fraction of particles with mass significantly different than the average is small, and the presence of these varied macroparticles has no significant effect on the dynamics.

To resolve the kinetic length scales of reconnection, which automatically resolves modes on the scale of the current sheet width, we set the grid spacing $\Delta x=\lambda_{\rm p}/8$, which ensures that the kinetic spatial and temporal scales are adequately resolved. The convergence tests by  \citet{kagan_13} discussed earlier in this section found that the evolution is insensitive to increases in resolution beyond this value, but we also carry out separate convergence tests that indicate similar physics up to $\Delta x=\lambda_{\rm p}/20$.

Finally, to ensure that we are able to probe nonlinear reconnection in a quasi-steady state, the size of the simulation must be much larger than the fastest-growing linear tearing mode wavelength of order $10 \delta$, and the duration $t_{\rm max}$ of the simulation must be significantly larger than the time required for an Alfv{\'e}n wave to cross the box (see \citet{kagan_13} for a detailed discussion).  We choose $L_x=800 \lambda_{\rm p}$, $L_y=600 \lambda_{\rm p}$, which is much larger than  the tearing length scale of $10 \delta=30 \lambdap$. { These large spatial scales for the box make adding 3D effects impractical because it would require larger computational resources by a factor of approximately 5000}.

For our simulations, in which $\sigma \gg 1$, the Alfv{\'e}n speed is close to the speed of light, so the duration of the simulations must satisfy $\omegap t_{\rm max}>800$, where $\omegap =c/\lambdap$ is the plasma oscillation frequency. We choose to run our simulations for a total duration of at least $\omegap t_{\rm max}=2500 $, which is significantly longer than this required duration. Energy is conserved within $0.5 \%$ in all of our simulations.

\subsection{Synchrotron radiation calculations}

In this paper, we calculate the radiation spectrum assuming that particles radiate solely through synchrotron radiation in the limit where radiative feedback is negligible;   Because the presence of electric fields means that the synchrotron radiation formulae cannot be used in the lab frame, and there may be no frame in which the electric field vanishes in a reconnection region we use the prescription of \citet{wallin_15} to calculate an effective magnetic field $B_{\rm eff}$ that would produce the instantaneous radius of curvature of the particle; this replaces all factors of $B \sin \alpha$, where $\alpha$ is the angle between the particle direction of motion and the magnetic field direction, in synchrotron formulae. Assuming that the force during a single tilmestep is adequately approximated by the Lorentz force at the particle's position, the effective magnetic field is 
\begin{equation}
B_{\rm eff}=\frac{m c \gamma}{q}\frac{\sqrt{p^2 F_L^2-({\mathbf p}\cdot \mathbf{F}_L)^2}}{p^2},
\end{equation}
where $\mathbf{F}_L$ is the instantaneous Lorentz force resulting from the electromagnetic field at the particle location.

Then the total synchrotron power $P$ is given by

\begin{equation}
P=\frac{2q^4\gamma^2 B_{\rm eff}^2}{3 m^2 c^5},
\end{equation}
and the synchrotron spectrum for each particle is given by 
\begin{equation}
\frac{dF_{\omega}}{d \omega}=\frac{\sqrt{3}q^3B_{\rm eff}}{2\pi m c^2} F\left(\frac{\omega}{\omega_c}\right),
\end{equation}
where $\omega$ is the radiation frequency and $\omega_c$ is the peak radiation frequency,
\begin{equation}
\omega_c=\frac{3 q B_{\rm eff}\gamma^2}{2 m c},
\label{eq:critomega}
\end{equation}
and $F$ is the synchrotron function
\begin{equation}
F(x)=x\int^\infty_{x}K_{5/3}(x) dx,
\end{equation}
where K represents a modified Bessel function of the second kind.

 In calculating the synchrotron spectrum as a function of direction, we do not use the fully general formula in which radiation at different frequencies has different angular distribution; instead, we assume that all radiation is distributed according to the angular distribution of total power per unit solid angle \citep{hoffman_2004}
 
 \begin{equation}
f(\theta)=\frac{7}{12}f_0 (1+(\theta \gamma)^2)^{-5/2}\left(1+\frac{5}{7(1+(\theta \gamma)^{-2})}\right)
\end{equation}
where  $\theta$ is the angle between the particle direction and the observer direction and $f_0$ is the power per unit solid angle of radiation emitted in the direction of particle motion $\theta=0$. We then use a kernel to allocate the total radiation in a discrete grid of observer directions. Our calculations of synchrotron power are in arbitrary units, and our calculated radiation frequencies are normalized to (within order unity) $0.29\omega_c$, so our synchrotron radiation calculations, like our particle-in-cell calculations, do not depend on the absolute value of the magnetic field $B$.

\section{Results} \label{sec:results}

\begin{figure}
\begin{center}
\includegraphics[width = 0.45\textwidth]{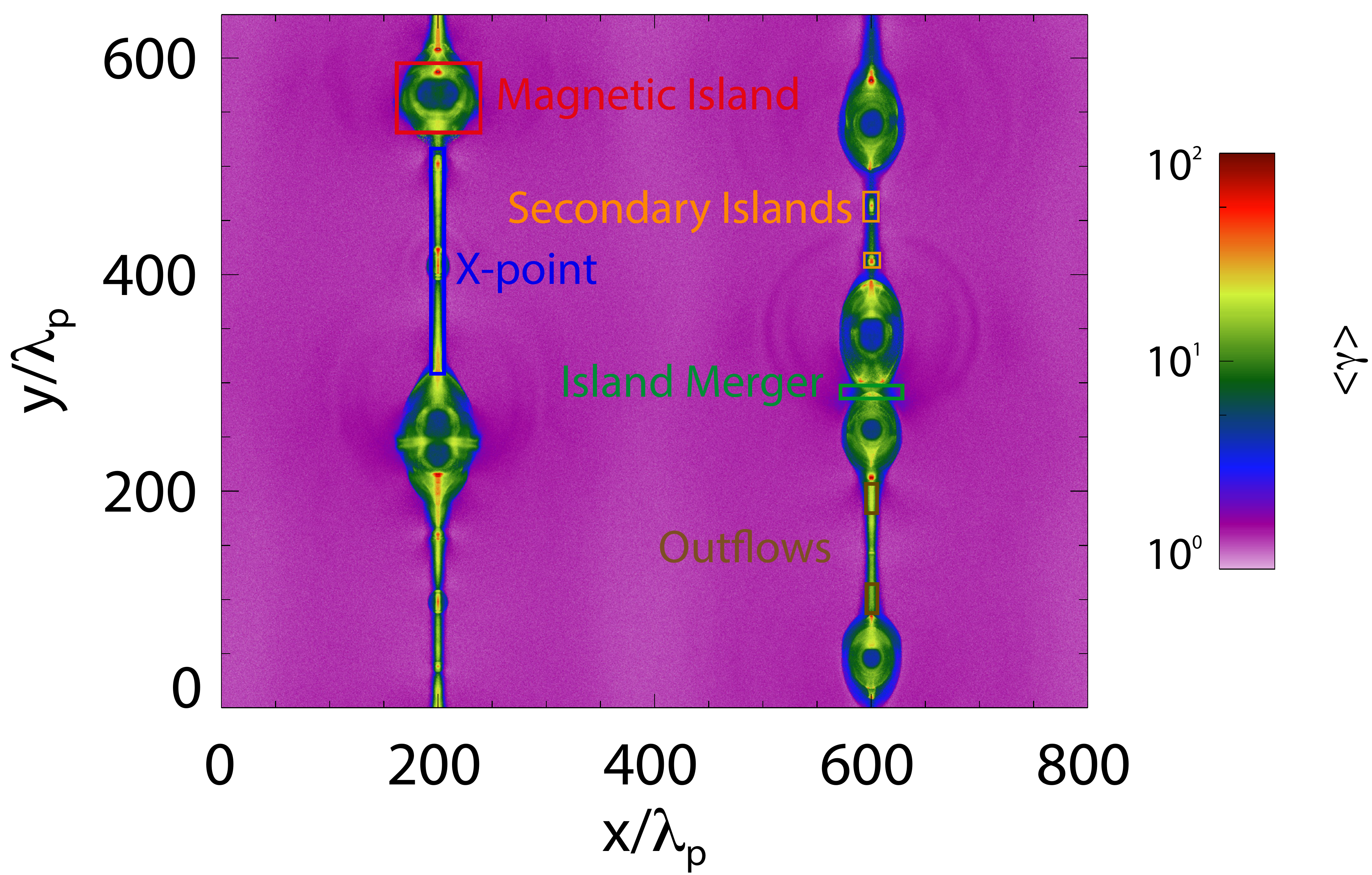}
\end{center}
\caption{The average kinetic energy as a function of location in simulation {\tt S4} in the middle of the reconnection process at $\omegap t=591 $. Boxes and labels of the same color  show the schematic features of reconnection: an X-point (blue),  a magnetic island (red), a group of secondary islands within an X-point (orange), a pair of merging islands (green), and outflows from an X-point towards two islands (brown).
\label{fig:schematic}}
\end{figure}

The  evolution of the current sheet in all of our simulations is similar to that found in previous 2D simulations of magnetic reconnection. The initial configuration is unstable to the tearing instability seeded by noise, and this instability grows producing chains of alternating X-points where reconnection occurs and dense magnetic islands where the outflows from reconnection meet. The instability becomes nonlinear at $\omegap t \sim 300$ in all simulations, and the islands begin to merge in hierarchical fashion until only one is left at typical time $\omegap t \sim 700$, at which point fast energy transfer stops. When the X-points and islands become large enough, secondary islands and secondary X-points in between them can form due to secondary tearing instabilities in the large X-points or in the current sheet produced between merging magnetic islands.  Figure \ref{fig:schematic} shows schematically these structural elements of the reconnecting current sheet in our simulation with $\sigma_0=4$ during this nonlinear stage. In the X-points the particles are accelerated by the electric field in the $\pm z$ direction. while at the edges of the X-point, particles flow towards the islands in the $\pm y$ direction as part of the outflow from the X-point. As we show later in Section \ref{sec:velmapping}, particles in the X-point tend to have most of their motion in this plane, with different species having opposite motions in the $z$ direction but similar motions in the $y$ directions at a given location.

 In this paper, we focus on the configuration of reconnection midway through this process at $\omegap t \sim 500$, a time when the rate of conversion of magnetic energy to kinetic energy is near its peak and the simulation is not yet affected by the boundary conditions { because there has not been time for for a light wave to cross the box}. In this way, we ensure that our results are not sensitive to the size of the simulations.  Because the initiation of reconnection occurs at slightly different rates in each simulation we choose a slightly different time in each simulation for detailed analysis.  Our results, which are discussed throughout this section, are summarised in Table \ref{tab:overview}.

\begin{deluxetable}{cccccc}
\scriptsize   
  \tablewidth{0pt}
\tablecolumns{6}

  \tablecaption{Table of Simulations \label{tab:overview}}
\tablehead{\colhead{Run\tablenotemark{a}} & \colhead{$r_{\rm rec, max}$\tablenotemark{b}}&\colhead{$\langle p_{y, \rm \ typ}\rangle/mc$ \tablenotemark{c}}&\colhead{ $\alpha$\tablenotemark{d}}&\colhead{$\gamma_p$\tablenotemark{e}}&\colhead{ Trough? \tablenotemark{f}}}
\startdata 
{\tt S4}&0.15&1.55&1.65&25&No\\
{\tt S40}&0.20&1.82& 1.5&130&Yes\\
{\tt S400}&0.17&2.04&1.4&180&Yes
\enddata
\tablenotetext{a}{ The number in each run indicates the initial value of $\sigma_0$ in that simulation.}
\tablenotetext{b}{$r_{\rm rec} $ is the reconnection rate calculated using the method discussed in Section \ref{sec:reconnectionrate}.}
\tablenotetext{c}{ The typical bulk momentum in the outflow direction, measured as described in the caption to Figure \ref{fig:bulkflow}. }
\tablenotetext{d}{The power law index of the high-energy tail of the energy spectrum for all particles initially located outside the current sheets.}
\tablenotetext{e}{ The approximate value of the maximum Lorentz factor of the power law at late times. The value for simulation {\tt S400} is affected significantly by boundary conditions, so it is probably not reliable.}
\tablenotetext{f}{Indicates whether the particle energy spectrum at the center of the current sheet in an X-point in the simulation has a trough separating a group of low energy-particles from another population of much higher-energy particle rather than a continuously decreasing power law.}
\end{deluxetable} 

 While many aspects of the simulations do not depend qualitatively on $\sigma_0$, significant differences are found between simulation {\tt S4} with $\sigma_0<10$  and simulations {\tt S40} and {\tt S400} with $\sigma_0>10$. In simulation {\tt S4}, the current sheets remain thin throughout the evolution, consistent with the results of \citet{2014ApJ...783L..21S}. As a result, the aspect ratios of the large-scale X-points are large, and secondary islands and X-points are produced within them by secondary tearing instabilities.  In contrast, for simulations {\tt S40} and {\tt S400} we find that the current sheet width increases as the structures become larger,  and aspect ratios remain small, of order 10. As a result, no substructure is formed in these current sheets. This is likely a result of the initial conditions in our simulations, in which density contrast is used in the initial conditions to maintain pressure balance.  \citet{bessho_12} found in their two-dimensional relativistic simulations that when the density contrast in the current sheet, which in our simulations with $T=0.5mc^2$ is approximately $5 \sigma_0$, is significantly larger than 10, a phase change in reconnection occurs which leads to widening of the current sheet and produces significantly different reconnection physics. In this paper, we examine what aspects of reconnection are significantly affected by this phase change. 

This section is organized as follows. In Section \ref{sec:overallevol}  we discuss the overall evolution of the reconnection configuration in the three simulations. In Section \ref{sec:reconnectionrate} we discuss methods of measurement of the reconnection rate in our simulations and compare our results to those found in other simulations.  In Section \ref{sec:partaccel} we investigate the particle energy spectrum produced by acceleration in the overall simulation and the local particle energy distribution at various locations in the current sheet. In Section \ref{sec:velmapping} we calculate the spatial distribution of velocities for particles of different energies throughout the current sheet. In Section \ref{sec:beaming}, we calculate the angular distributions of particle velocities and the resulting synchrotron radiation, and investigate whether strong beaming of either is produced in our simulation.

\subsection{Overall evolution}\label{sec:overallevol}
\begin{figure*}
\begin{center}
\includegraphics[width = \textwidth]{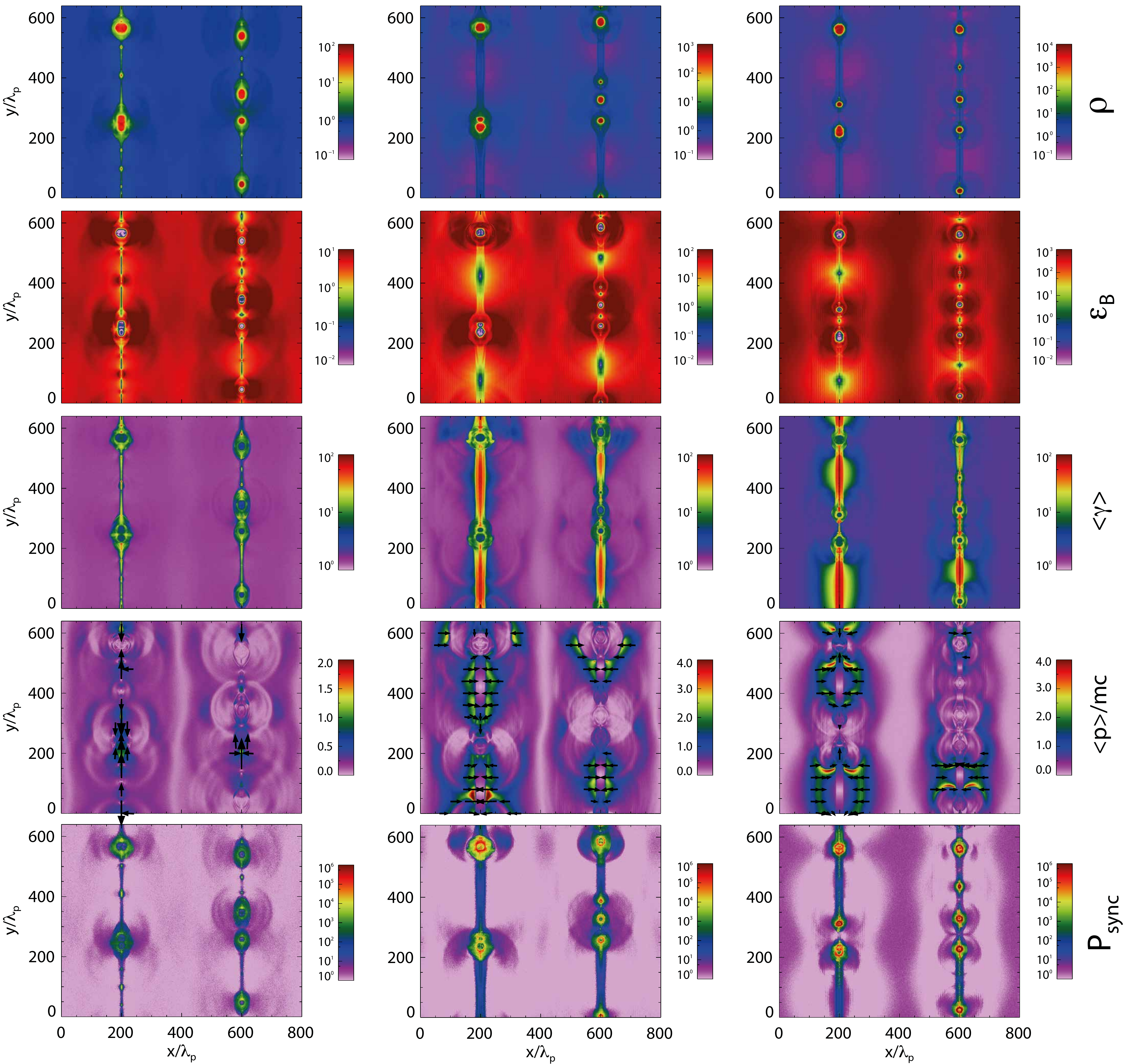}
\end{center}
\caption{ The structure of the simulation with $\sigma_0=4$ at time $\omegap t=591 $ (left column) $\sigma_0=40$ at time $\omegap t=534 $ (left column), and $\sigma_0=400$ at time $\omegap t=478$ (right column). The colors show the total number density $n$ normalised to the background density $n_b$ (top row),    Magnetic energy to total particle rest mass ratio  $\epsilon_b=B^2/8 \pi n m c^2$ (second row), average particle kinetic energy $\langle \gamma \rangle$ (third row), the bulk momentum of the plasma $\langle p \rangle/mc$ (fourth row), and the total synchrotron power emitted in arbitrary units (bottom row). The direction of the arrows in the fourth row indicates the direction of the bulk momentum. The size of the arrowheads and lengths of the arrows are proportional to the bulk momentum in each panel, but their sizes vary between panels to clearly illustrate the direction of the bulk momentum. Arrows are only shown for locations with $\langle p \rangle/mc>0.5 $ for $\sigma_0=4$ and  $\langle p \rangle/mc>2.0 $ for $\sigma_0=40$ and $\sigma_0=400$. \label{fig:overallevol}}
\end{figure*}

Figure \ref{fig:overallevol} shows the reconnection configuration for $\sigma_0=4$ at time $\omegap t=591 $ (left column), $\sigma_0=40$ at time $\omegap t=534 $ (middle column), and $\sigma_0=400$ at time $\omegap t=478$ (right column).  The top three rows of panels show the basic characteristics of relativistic reconnection in these simulations. The tearing instability has produced alternating X-points and magnetic islands, with particles entering the current sheet at X-points and flowing out into the magnetic islands flanking the X-points in both directions. Secondary current sheet structures are identifiable at the locations of island mergers. Although the limited island size at the chosen time of investigation makes the detailed structure of these current sheets unclear, examination of similar secondary current sheets at later times indicates that they have the same X-point  and magnetic island structures found in the primary current sheets.  
 The plot of density $\rho$ in the first row of Figure \ref{fig:overallevol} shows that in all of our simulations most of particles that enter the current sheet are eventually concentrated in the centers of magnetic islands. However, in contrast to the simulations of \citet{2014ApJ...783L..21S}, we find that these particles are not highly accelerated. This difference likely arises from the differing initial conditions; in our simulations, all particles begin at the same temperature, while in those of \citet{2014ApJ...783L..21S}, the current sheet population begins with Lorentz factors similar to $\sigma_0$.  Particles in the center of the  islands are likely to be those that initially entered the X-points; such particles have had time to isotropize and are more likely to remain in the center of the islands. Particles with high energies that enter the magnetic islands at later times go into the current sheet with a significant initial momentum in the $x$ direction and therefore oscillate across the islands, spending most of the time at island edges where their direction reverses. Therefore, the different temperatures found at island centers are a result of the initial conditions, rather than differing reconnection physics. The second row of Figure \ref{fig:overallevol}, which shows the magnetic energy to particle rest mass ratio $\epsilon_B$, confirms that both the centers of islands and the X-point regions are dominated by particle energy as expected in magnetic reconnection.  

Detailed examination of the third row of Figure \ref{fig:overallevol} indicates that particles entering the current sheet at X-points where reconnection is taking place are initially mildly relativistic with $\gamma\sim 2$, but are accelerated to a typical Lorentz factor larger than 10 in the reconnection region.  Conservation of energy implies that the typical Lorentz factor in the current sheet measured in the lab frame should be approximately equal to $\sigma_0$ if the background particles are cold.  For simulation {\tt S40} with $\sigma_0=40$, and simulation {\tt S400} with $\sigma_0=400$, the results shown in the third row of  Figure \ref{fig:overallevol} are broadly consistent with this expectation (note that the values of $\langle \gamma\rangle$ in the center of the current sheet for simulation {\tt S400} can be as large as $200$). In contrast, for simulation {\tt S4}, the typical Lorentz factor in the current sheet is significantly larger than $\sigma_0=4$,  $\langle \gamma\rangle\sim 20$, but this is consistent with the maximum $\gamma_p$ of the power law discussed in the next section.  
\begin{figure}
\begin{center}
\includegraphics[width = 0.45\textwidth]{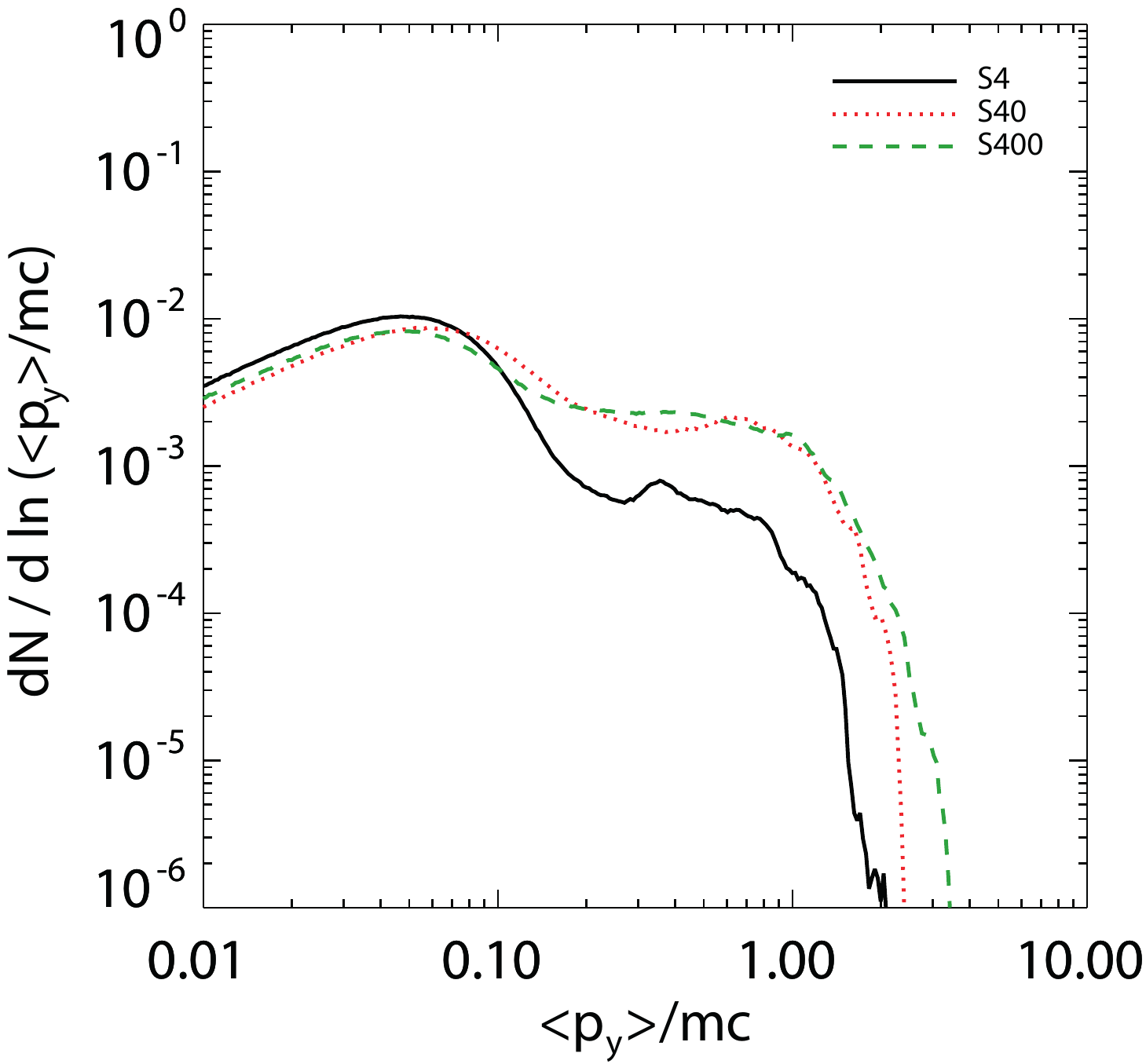}
\end{center}
\caption{The distribution of the bulk outflow momentum $\langle p_{y}\rangle/mc$ calculated at locations with $x$-coordinate within $\pm 10\lambda_p$ of a current sheet for each run. The distribution is flat up to approximately  $\langle p_{y}\rangle/mc=1$ in all simulations, and then declines quickly. We calculate a typical outflow bulk momentum $\langle p_{y, \rm \  typ}\rangle/mc$ (shown in Table 1) by finding the value of $\langle p_{y}\rangle/mc$ where the distribution falls by a factor of 10 from its value at $\langle p_{y}\rangle/mc=1$. This parameter does not depend strongly on the choice of cutoff or the width of the current sheet region at which we calculate $\langle p_{y}\rangle/mc$. Distributions are normalized so that they sum to 1. \label{fig:bulkflow}}
\end{figure}

The fourth row of Figure \ref{fig:overallevol} shows the bulk momenta of the plasma in the simulations. It indicates that there are three locations in which significant bulk flows are present. The first location is the inflow region of X-points, which has only mildly relativistic flows of approximately $\langle p_x\rangle/mc\sim0.2$ in simulation {\tt S4} but highly relativistic inflows with $\langle p_{x}\rangle/mc>2$ in simulations {\tt S40} and {\tt S400}; these high velocities are consistent with those found by \citet{bessho_12} in their simulations of low-density plasmas. However, these flows are intermittent and we show later that they do not imply an extremely high rate of energy transfer overall. 

The second location of significant flow is in apparent shock waves produced by the collisions of magnetic islands, which produce disturbances in the background material that increase in importance with $\sigma_0$. However, these disturbances are not associated with significant energy transfer or radiation.  It should be noted that individual species of particles tend to have oppositely oriented fast flows in the $\pm z$ direction in the center of the X-point which are not shown in calculations of bulk motions; we discuss these in section \ref{sec:velmapping}. 

The third location of significant velocity flow is the outflow regions of X-points as particles move towards magnetic islands. These locations are are typically thought to contain the fastest flows produced in relativistic reconnection, with typical flow momenta of $\sim \sqrt{\sigma_0}$ found in the simple model of  \citet{2005MNRAS.358..113L} as well as the simulations of \citet{2014ApJ...783L..21S} and \citet{sironi_16}. {Figure \ref{fig:bulkflow} shows the outflow bulk momentum distribution in the current sheet. In all runs, this distribution is flat up to approximately $\langle p_{y}\rangle=mc$, and that the cutoff above this value does not depend strongly on $\sigma_0$. Specifically, we find that the typical bulk momentum in the outflow direction $\langle p_{y, \rm \ typ} \rangle /mc $ defined in the caption to Figure 3 increases only slowly with $\sigma_0$ from $1.55-2.04$ (see Table 1), which in the highest-magnetisation case of simulation {\tt S400} is much smaller than the prediction of $\sqrt{400}=20$. This result is resolution-independent, varying by at most $9\%$ with macroparticle density and cells/skin depth for all values of $\sigma_0$,} and it is consistent with the finding by \citet{melzani_14} in the case of electron-ion plasmas that the vast majority of the kinetic energy produced in high-$\sigma$ reconnection is converted to random motion, rather than ordered bulk outflows. 

Finally, the fifth row of Figure \ref{fig:overallevol} shows the total synchrotron power emitted at each location in arbitrary units. The regions of significant emission are the outflow regions of X-points and the outlying areas of islands, with particularly high emission located at the points where small islands meet larger islands, such as the location $x=200 \lambdap$, $y=580\lambdap$ in simulation S4. Unlike the case of the bulk momenta, contributions to the synchrotron radiation from the inflow region of X-points in simulations {\tt S40} and {\tt S400} are relatively small, indicating that any beaming that may occur there does not have a strong effect on the radiated emission.   This indicates that that the locations of synchrotron emission do not depend qualitatively on the value of $\sigma_0$. The oppositely oriented flows of individual species at the center of X-points also have little effect on the overall emission, because they are located in a region of small magnetic field. In our later calculations of synchrotron emission and beaming, we focus on simulation {\tt S40} in our analysis but confirm that there is no significant qualitative dependence on $\sigma_0$ in our results.

\subsection{Reconnection Rate}
\label{sec:reconnectionrate}
The rate of conversion of magnetic energy to kinetic energy in reconnection is of paramount importance for the fast production of possibly beamed radiation in relativistic reconnection sites. In general, reconnection rates are defined by assuming a steady-state reconnection equilibrium with inflow velocity $v_{\rm in}$ and outflow velocity $v_{\rm out}$. Then the normalised reconnection rate is given by  $r_{\rm rec}=v_{\rm in}/v_{\rm out}$, which implicitly assumes that the efficiency of conversion of magnetic to kinetic energy is very high once plasma enters the current sheet. This assumption is indeed justified by the fact that the current sheet is dominated by the energy of the particles in numerical simulations, as shown in the the plot of $\epsilon_B$ in Figure \ref{fig:overallevol}, but this is not a direct measurement.  Due to the difficulty involved in calculating the typical velocity of inflow, which varies strongly with position, one instead typically measures the nearly uniform electric field $E_y\approx v_{\rm in} B_0/c$, which gives a normalised reconnection rate of \citep{bessho_12, melzani_14}

\begin{equation}
r_{\rm rec}=\frac{E_y}{v_{\rm out} B_0}=\frac{E_y}{v_{\rm A,in} B_0},
\end{equation}
where the second part of the equation  reflects the common assumption that the outflow speed is equal to the Alfv{\'e}n speed in the inflow region $v_{\rm A ,in}$ which is given in the relativistic case by \citep{gedalin_93}
\begin{equation}
v_{\rm A,in} =\frac{c}{\sqrt{1+\sigma_0^{-1}}}.
\end{equation}
This corresponds to an outflow bulk momentum of

\begin{equation}
p_{\rm out}/mc= \sqrt{1+\sigma_0},
\end{equation}
which is consistent with the predictions of  \citet{2005MNRAS.358..113L} and the simulations of \citet{2014ApJ...783L..21S}  which have reconnection outflow Lorentz factors $\sim \sqrt{\sigma_0}$ but not with our simulations or those of \citet{melzani_14}, which find moderate outflow Lorentz factors of $\sim 3$ which do not depend strongly on $\sigma_0$.   A somewhat more complex local definition for reconnection rate has been proposed by \citet{liu_15}, which similarly predicts high outflow Lorentz factors at high $\sigma_0$. The reconnection rate calculated will not be strongly affected by this disagreement because $v_{\rm A, in}$ does not depend strongly on $\sigma_0$, but the low values of $p_{\rm out}$ found in some simulations cast doubt on the physical realism of the models.

 In addition to this shortcoming, the local reconnection rate may not be a good proxy for the global rate of energy transfer because inflows only occur at X-points rather than throughout the whole current sheet. In the case of low-density plasma, \citet{bessho_12} find that the local reconnection approaches 1 at high $\sigma_0$ while remaining close to the typical value of $0.05-0.2$ found in most simulations at low $\sigma_0$, and our results are similar. But the normalised global rate of energy transfer (discussed below) is similar at all values of $\sigma_0$ in our simulations, so the local reconnection rate is not an adequate measure of the global energy transformation rate.
 
\citet{2014ApJ...783L..21S} define a global measure of reconnection rate by choosing $v_{\rm in}$ as the average velocity of inflow far from the current sheet. However,  this definition is somewhat difficult to apply because the distance from the current sheet must be chosen in a somewhat arbitrary manner and flows resulting from magnetic island collisions far from the current sheet may bias the result.  Instead, we propose a definition of reconnection rate based directly on the global characteristics of energy transfer. The maximum rate at which energy transfer can take place by reconnection occurs when at all locations along the current sheet the inflow speed is $c$ and all magnetic energy is converted to kinetic energy. The energy gained in this case accounting for the two directions of inflow into each current sheet and the two current sheets present in the simulation is approximately

\begin{equation}
\frac{d \mathcal{E}_{\rm K,max}}{dt}=4 L_y c \frac{B_0^2}{8\pi}=\mathcal{E}_{\rm B,0} \frac{4 c}{L_x},
\end{equation}
where $\mathcal{E}_{\rm B,0}=L_x L_y B_0^2/8 \pi $ is the total energy in the magnetic field at the beginning of the simulation. Then, we can define a dimensionless reconnection rate by normalising the rate of change of total kinetic energy to this value, yielding

\begin{equation}
r_{\rm rec}=\frac {d \mathcal{E}_{\rm K}}{dt}\frac{L_x}{4 c \mathcal{E}_{\rm B,0}}.
\end{equation}
The generalization of this definition to three dimensions yields the same equation.

We find that the reconnection rate calculated in this way does not vary strongly with time during the period of fast reconnection. Table 1 shows the average value of this reconnection rate in each simulation from the beginning of nonlinear reconnection at time $\omegap t=340$ until the end of the reconnection phase, chosen by the time when significant energy transfer stops. Using this definition, we find that the reconnection rate does not depend strongly on $\sigma_0$, but instead occurs at a universal rate of rate $r_{\rm rec}\sim0.15-0.2$ which is approximately consistent with the results of past two-dimensional simulations of relativistic reconnection at lower magnetisation \citep[e.g.][]{liu_11,bessho_12,melzani_14,liu_15}.   We find that despite the significant differences in the current sheet thickness and the relativistic inflow speeds found in our simulations at different $\sigma_0$, the global reconnection rate does not vary greatly.  {This conclusion needs testing in 3D simulations, in which the reconnection rate can be significantly smaller \citep{2014ApJ...783L..21S}, but if it holds in that case our finding that the reconnection rate is independent of $\sigma_0$ in relativistic reconnection will be confirmed.}

\subsection{Particle energy spectra and their variation }
\label{sec:partaccel}

We now consider the particle energy spectrum resulting from relativistic reconnection. In all simulations, reconnection produces strong particle acceleration, and at around $\omegap t=500$ approximately $30\%$ of the magnetic energy in each simulation has been converted to kinetic energy. The resulting spectrum for all simulations is a Maxwellian plus a power law, with a cutoff at high energy. Table 1 shows that the power law becomes harder at larger values of $\sigma_0$, with the index $\alpha$ of the power law spectrum $dN/d \gamma \propto \gamma^{-\alpha}$  changing from 1.65 at $\sigma_0=4$ to 1.3 at $ \sigma_0=400$. This relation is in approximate agreement with the results of previous simulations \citep{2014ApJ...783L..21S,guo_14,guo_15} as well as the detailed study of \citet{werner_16}.

The hard power laws found in our simulations indicate that the maximum $\gamma$ should reach saturation at late times as a result of energy conservation. The maximum possible value of the average Lorentz factor of the particles is $\bar{\gamma} m n_b c^2 \le B_0^2/8\pi$, which yields $\bar{\gamma}= f\sigma_0 h$, where $f$, which is expected to be of order unity, is the fraction of the magnetic energy that is converted into kinetic energy when a part of the plasma undergoes reconnection. If all of the particles are located in a power law with $1<\alpha <2$ beginning at $\gamma_0$ and ending at $\gamma_p$, with $\gamma_p \gg \gamma_0$, the average Lorentz factor is

\begin{equation}
\bar{\gamma}=f \gamma_0 \frac{2-\alpha}{\alpha-1}\left(\frac{\gamma_p}{\gamma_0} \right)^{2-\alpha}.
\end{equation}

Then, we have 

\begin{equation}
\gamma_p=\gamma_0 \left(\frac{\sigma_0 h}{f \gamma_0}\frac{\alpha-1}{2-\alpha} \right)^{1/(2-\alpha)} .
\label{eq:gammap}
\end{equation}

For $h=2.5$ as applicable to our simulations, $\alpha=1.5$, and $\gamma_0=5$, which is consistent with the measured particle energy spectra at all values of $\sigma_0$ , we find that as an upper limit, $\gamma_p \sim \sigma_0^2$. \citet{werner_16} found in their simulations that the actual saturation value is close to $\sim 4 \sigma_0$. We find that the power law spectra in our simulations  indeed reach saturation at late times at Lorentz factors shown in Table 1, and that the maximum Lorentz factor at saturation $\gamma_p$ increases with $\sigma_0$. For $\sigma_0=4$, the value of $\gamma_p=25$ is comparable to both predictions (which are similar),   For $\sigma_0=40$, the value of $\gamma_p$ we find is compatible with the predictions of \citet{werner_16}. And finally, for $\sigma_0=400$, the value of $\gamma_p$ is much smaller than either prediction. These results may be explained by the fact that $\gamma_p$ is also limited by the total size and timescale of the simulation.  Specifically, the maximum possible energy gain for a particle accelerated over a time $t$ is given by $\Delta \gamma=e E t/m_e c^2$, which may be combined with the definitions of $\sigma_0$ and $\omegap$ to yield $\Delta \gamma \sim (E/B_0)\omega_p t$; the local value of $E/B_0\sim 0.5$ in our low density simulations {\tt S40} and {\tt S400}. Because fast reconnection in our simulations extends over a maximum period of $ \omega_p t \sim400$, it is clear that the simulation size is limiting acceleration for simulation {\tt S400}, but not for the other simulations.

Physically, the fact that $\gamma_p$ is significantly smaller than predicted  by Equation \ref{eq:gammap} is an indication that particles in the background of the simulation and those originally located in the current sheet have received some of the dissipated magnetic energy. The plot of $\langle \gamma\rangle$ in Figure 2 indicates that in simulation {\tt S400}, the whole background plasma has been heated by the reconnection in the current sheet from the initial value of $\langle \gamma\rangle\sim 2$  at $\omegap t =0$ to  $\langle \gamma\rangle\sim 4$ at $\omegap t = 478$. This strong heating is likely a result of the boundary conditions becoming important in this simulation. For simulation {\tt S40}, the disagreement of the value of $\gamma_p$ with the prediction of Equation (\ref{eq:gammap}) arises because most ($\sim 83 \%$) of the energy converted is transferred to the particles initially located in the current sheet; thus, less energy is provided to the inflowing particles and they do not reach high $\gamma_p$.

We now investigate in detail the particle energy spectra for simulation {\tt S40} with $\sigma_0=40$, including an investigation of its variation with location. The only significant qualitative difference between the spectra at different values of $\sigma_0$ is the presence of a "trough" in the X-point spectrum at intermediate energy for $\sigma_0=40$ and $\sigma_0=400$ and the absence of such a trough for $\sigma_0=4$, which will be discussed later in this section. This qualitative difference is present in the simulations of \citet{bessho_12} for low-density plasmas that correspond to high $\sigma_0$, as well. 

\begin{figure}
\begin{center}
\includegraphics[width = 0.48\textwidth]{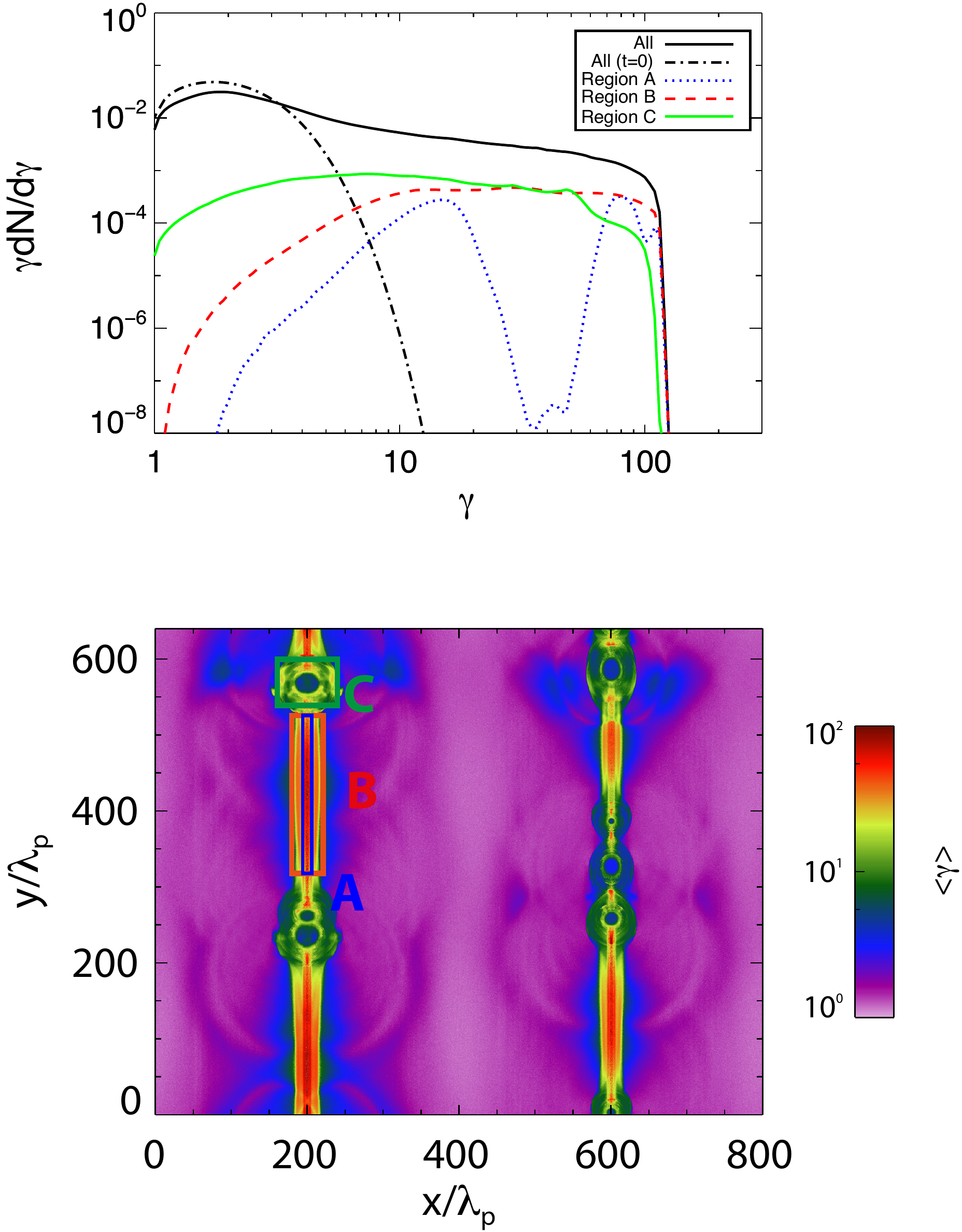}
\end{center}
\caption{ (top panel) The energy spectra $\gamma dN/d\gamma$ of the positrons in simulation {\tt S40}  normalised to the total particle number for the whole simulation box (black), the center of the current sheet in an X-point at $196<x/\lambdap<204$, $320<y/\lambdap<520$ (Region A, red), the whole X-point at $184<x/\lambdap<216$, $320<y/\lambdap<520$ (Region B, blue), and the magnetic island at $160<x/\lambdap<240$, $540<y/\lambdap<600$ (Region C, green). The initial energy spectrum for the whole box is also shown (black, dot-dashed). (bottom panel) The average kinetic energy $\langle \gamma \rangle$ at each location in the simulation, with boxes highlighting the locations of each region at which spectra are calculated.
\label{fig:partspec}}
 \end{figure}

In this analysis, and in most of our later results, we focus on the characteristics of the positively charged (macro)particles in order to display the differing motions of each species that produce the antiparallel flows that constitute the current in each current sheet. Results for the negatively charged particles are always very similar to the results for positively-charged particles, except that motions in the $\pm z$ directions are reversed for a given current sheet. Since we are simulating a pair plasma, we occasionally refer to the positively charged particles as positrons. The top panel of Figure \ref{fig:partspec} shows energy spectra in the overall box and in various regions of the box, while  the bottom panel of the figure shows several regions of interest for which we take particle energy spectra, which correspond to the center of the current sheet in an X-point (Region A), the entire X-point (Region B), and a magnetic island (Region C).

 The top panel of Figure \ref{fig:partspec} shows that the overall energy spectrum has a relatively weak thermal peak at $\gamma\sim 2$ with a hard power law tail that contains a large proportion of the total energy. The initial energy spectrum at the beginning of the simulation is also shown, illustrating that most particles with $\gamma>4$ have been significantly accelerated during the simulation. Region A, which corresponds to the center of the current sheet in an X-point, has a peculiar spectrum, consisting of a moderately accelerated population at $\gamma\sim 15$, which may correspond to the initial acceleration of particles flowing in to the current sheet, and another population at very high energies close to the maximum found in the whole box, with almost no particles found in the trough at intermediate energies. There is another trough present at $\gamma\sim 100$, indicating that multiple features of this type exist. This spectrum may be understood by considering that the particles being accelerated in an X-point are moving in Speiser orbits \citep{speiser_65} in which they oscillate across the current sheet while being accelerated by the electric field in the X-point. Because the initial inflow is relativistic, the particles that flow into the current sheet at a given time tend to have similar and large momenta in the $x$ direction with low random velocity and the particles will undergo few oscillations before leaving the X-point. Thus, particles in the X-point with a given value of $\gamma$ will all tend to be at a similar distance $|x-200\lambdap |$ from the center of the X-point. If the distance is high, a trough will be present and the particles will tend to have little momentum in the $x$ direction, while if the distance is low the trough will be absent and the particles will have large momenta in the $\pm x$ direction. Over longer timescales in which many Speiser oscillations can occur, we expect these features to disappear as the particles no longer move coherently. The spectrum in Region B indicates that when a larger portion of the current sheet is included, the trough disappears because the particles with high oscillation amplitude are now included in the spectrum.

The spectrum in the magnetic island (Region C) is very different than that found in the X-points. Instead of having a thermal peak at the background average energy of $\gamma \sim 2$ , the peak is at a significantly larger value of $\gamma\sim 7$, which is similar to the average kinetic energy in the center of the island where the density is extremely high.  Above this peak, the particle energy spectrum declines faster than the overall power law of $p=1.5$, but bumps are present at  $\gamma \sim 30$  and at $\gamma\sim 50$  which correspond to locations in Region C with similar values of $\langle \gamma \rangle$, corresponding approximately to the colors yellow and red in the bottom panel of Figure \ref{fig:partspec}.  The magnetic island appears to have relatively few particles with $\gamma>60$, indicating that the highest-energy particles are found mainly in the X-points. As long as fast reconnection is still taking place and the highest-energy particles have not reached saturation, the particles remaining in the X-points will always have a higher energy than those in the islands because they have spent a longer time being accelerated.
Overall, these spectra indicate that most particle acceleration takes place in reconnection regions, in which hard power laws are produced. The magnetic islands mostly contain particles that have been only slightly accelerated in the X-points, which are in an approximately thermal distribution peaked at $\gamma\sim 7$, and high-energy particles contribute less of the total particle energy in these regions.  The overall energy spectrum of particles that have entered the current sheet and been accelerated will be some combination of the spectra shown for Regions B and C. This spectrum peaks at around $\gamma\sim 10$, which is around $5$ times the initial average particle energy,  and has a hard power law spectrum that extends to high energies; the exact index of the power law depends on whether more particles are located in the X-points or the islands. We expect these accelerated particles to dominate the high-energy synchrotron emission, so we study in detail the particle distribution and the resulting radiation from these two regions in the subsequent sections.

 At late times in the simulation, as energy conversion through reconnection becomes slower, the islands become the dominant reservoir of the highest-energy particles due to either saturation of $\gamma_p$ or the influence of the boundary conditions, with most of these particles being located where outflows from X-points meet the outer layers of the island.  Nevertheless, particles in the X-points still have a higher average energy than those in the islands do at late times.  Because this saturation may be a result of artificial boundary conditions, we assume that the particle distributions at the earlier time focused on in this work are a more accurate representation of particle energy spectra in large-scale reconnection. 

\subsection{Beaming of particles as a function of location}\label{sec:velmapping}
  \begin{figure}
  \begin{center}
\includegraphics[width = 0.45\textwidth]{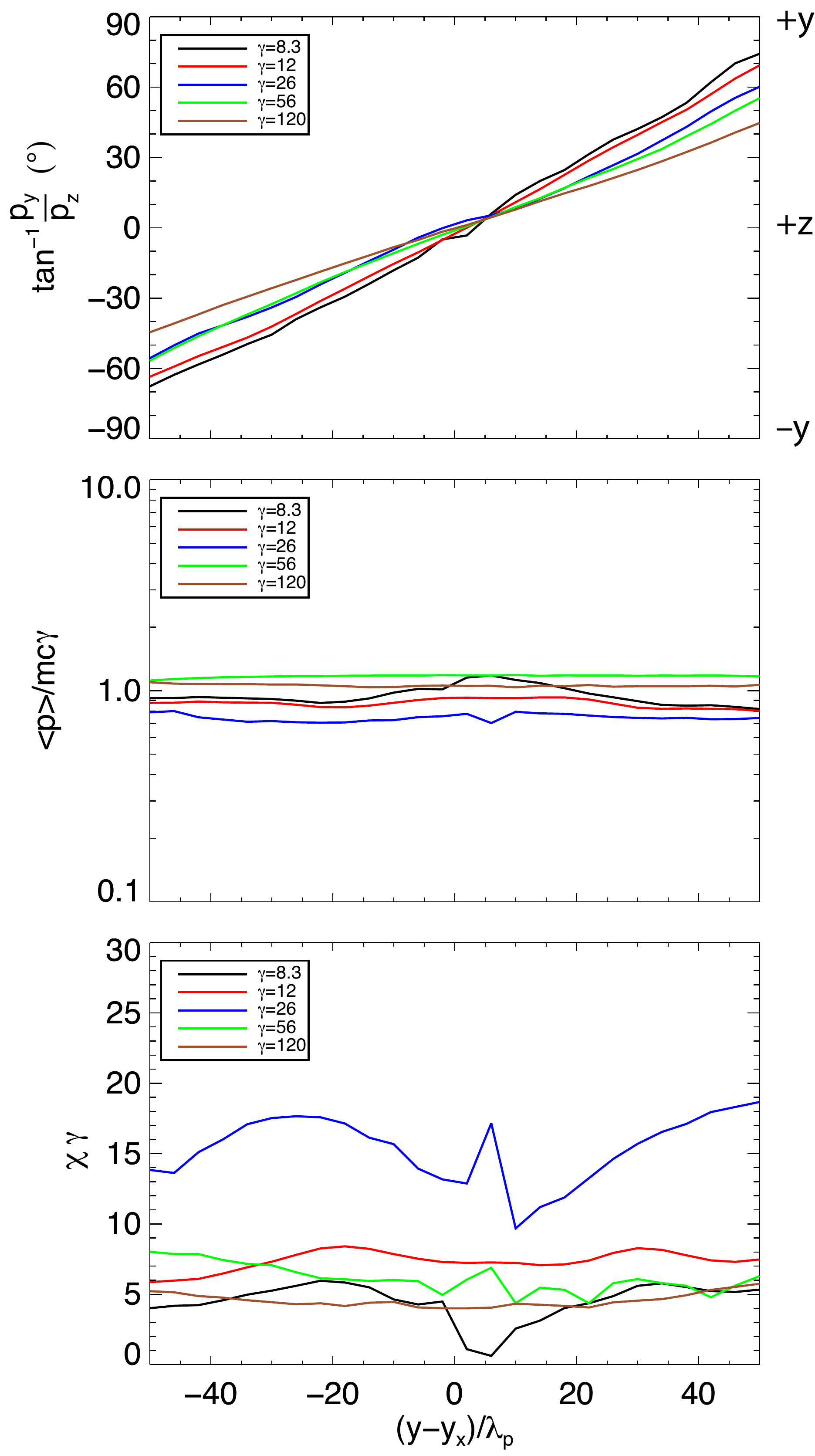}
\end{center}
\caption{ Characteristics of the velocity distribution for the positrons with Lorentz factors in energy bands centered on $\gamma=8.3,\ 12,\ 26,\ 56$, and 120 in Region A at $196<x/\lambdap<204$, $320<y/\lambdap<520$,  corresponding to the center of an X-point in the $x$ direction. The figure shows the angular direction of the bulk momentum in the plane of the current sheet (top),  bulk momentum normalised to the energy band $\langle p \rangle/mc \gamma$ (middle), and the beaming angle of the particle distribution normalised to $1/\gamma$ (details of the calculation are shown in the text). Each point represents a calculation of these quantities for particles  within a square of size $8\lambdap$ centered on the value of $y$ shown and the x-coordinate $x/\lambdap=200$ which corresponds to the center of the left current sheet. The values of $y$ are restricted to the range  $-100<(y-y_x)/\lambdap<100$, where $y_x=420 \lambda_p$ corresponds to the center of the X-point in the left current sheet. 
\label{fig:velmapping}}
 \end{figure}

Although we have not found bulk beaming in our simulations, it is still possible that strong energy-dependent kinetic beaming \citep{cerutti_12b} will be present. In this section, we discuss the average momentum and the spread in the direction of the momentum (i.e., the beaming) as a function of energy for particles at various locations in the simulation. We exclude particles that began in the current sheet from our calculations, just as in Section \ref{sec:partaccel}. Particles that are not significantly accelerated can have strong momentum flows in three locations, as explained in Section \ref{sec:overallevol}. However, particles that are significantly accelerated with $\gamma>5$ only have significant bulk momentum, and significant beaming, in the X-point and in the outflow regions from those X-points in the current sheet. Although bulk flows are not present in magnetic islands. we investigate whether there is beaming of high-energy particles in these locations, because they dominate the bolometric synchrotron flux.

\subsubsection{Beaming of particles in X-points}
We first study the beaming of particles in the X-points and their adjacent outflow regions.  We focus on the beaming of particles in Region A to include in the analysis the spread in the $x$ direction resulting from particles entering the X-point in the opposing $\pm x$ directions. In the X-point outside of this region, there is a net momentum in the $\pm x$ direction and very little spread in that direction, which is unrepresentative of the particle distribution in the X-points. Figure \ref{fig:velmapping} shows the characteristics of the momentum distributions of the positrons (positively charged macroparticles) as a function of $\gamma$ for particles located in the middle X-point in the left current sheet in Region A. The top panel of the figure shows the direction of bulk momentum in the $y-z$ plane; the average momentum of particles in the $x$ direction is close to 0 everywhere in Region A because the inflows are of similar strength on both sides of the X-point. It indicates that the direction of the bulk momentum changes gradually from the direction of the acceleration provided by the electric field ($+z$) towards the direction of the outflow ($\pm y$)  as one moves from the center of the X-point towards the magnetic islands. Positrons with higher energy are deflected somewhat less towards the outflow direction as they move away from the center of the X-point. This makes sense because such particles have significantly higher inertia following acceleration than lower energy particles do.

  The second panel of Figure \ref{fig:velmapping} shows the ratio of the bulk momentum of particles falling within a given energy band to the central Lorentz factor of that band. At all locations, the bulk momentum is similar to the total Lorentz factor, indicating that almost all of the particles are moving in the same direction at a given location. Note that the points with $\langle p \rangle/m c \gamma>1$ indicate that much of the bulk momentum is in the higher-energy particles within the band, and as a result the maximum possible bulk momentum is closer to $\langle p\rangle/m c \gamma_{\rm max}$ where $\gamma_{\rm max}$ is the maximum value of the Lorentz factor in the energy band. In all cases,$\langle p\rangle/m c \gamma_{\rm max}<1$, so this high degree of alignment is indeed physically possible.

 The third panel of Figure \ref{fig:velmapping} shows the ratio between the standard deviation $\chi$ of the direction of positron momentum to $1/\gamma$, the minimum opening angle of radiation arising from a distribution of particles with Lorentz factor $\gamma$.  We discuss this calculation in an appendix.  This panel shows that accelerated particles in the X-point typically are beamed within $\chi \gamma \sim 5-10$, with no significant dependence of the normalised value on $\gamma$.  A somewhat higher spread is found for the case $\gamma\sim 26$, which also has a significantly lower normalized bulk momentum in the middle panel of Figure \ref{fig:velmapping}, but this is a discrepancy of only a factor of $\sim 3$. The increased spread may occur because there are few particles in Region A at this value of $\gamma$, which is in the trough discussed in the previous section. These results indicate that accelerated particles located in the reconnection region are highly beamed, but that the direction of that beaming changes greatly as one moves from the center of the X-point to its edge. Thus, the overall particle distribution and the resulting energy distribution should be strongly kinetically beamed in the plane of the current sheet (the $y-z$ plane), but not in other directions.

\subsubsection{Beaming of particles in magnetic islands}
  \begin{figure}[h]
\begin{center}
\includegraphics[width = 0.45\textwidth]{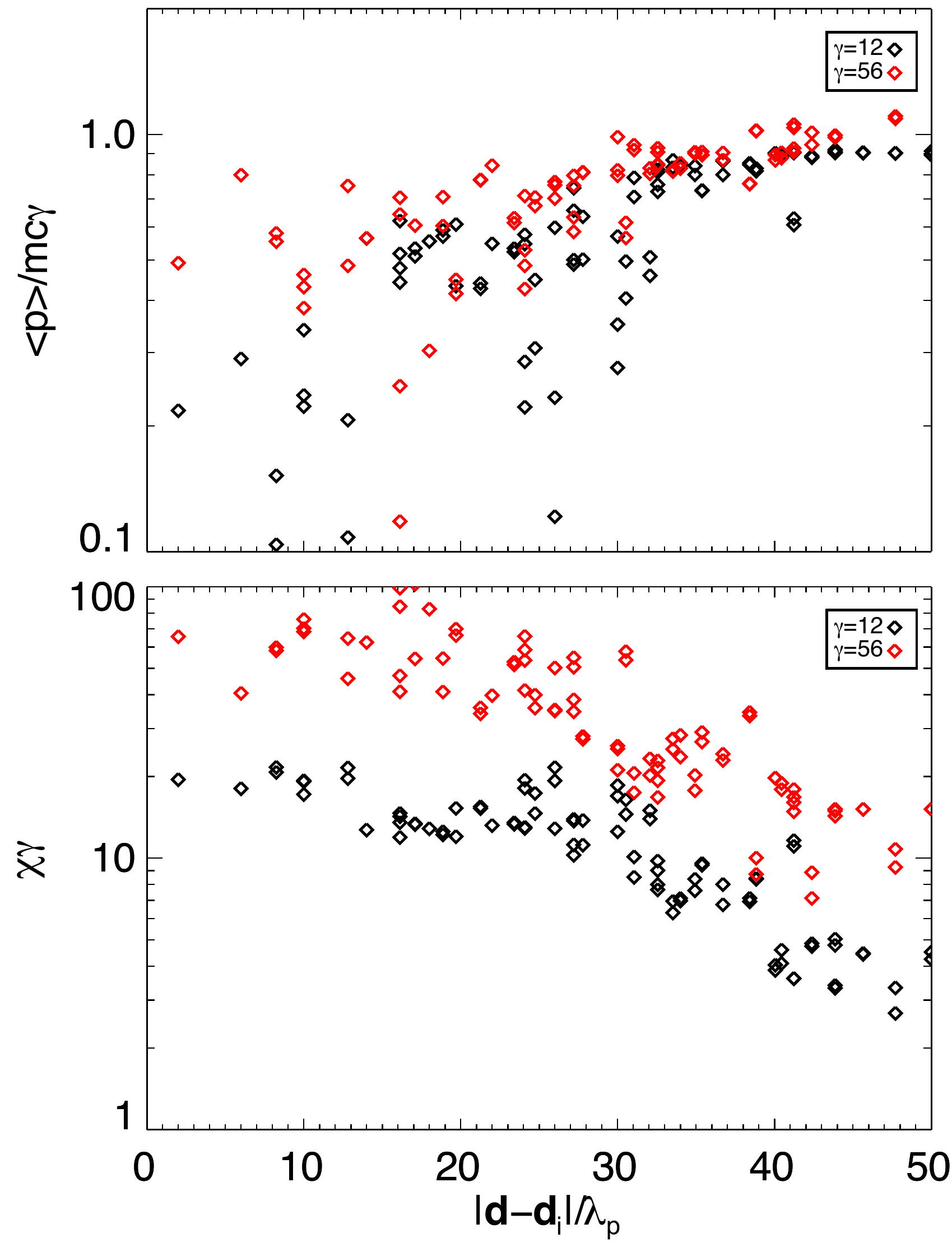}
\end{center}
\caption{ Characteristics of the velocity distribution for the positrons with Lorentz factors in energy bands centered on $\gamma=12$ and $\gamma=56$ in Region C corresponding to a magnetic island as a function of the distance $|\mathbf{d}-\mathbf{d}_i$, where $\mathbf{d}=(x,y)$ is a position vector, and  $\mathbf{d}_i=(200\lambdap,\ 570\lambdap)$ is the location of the center of the island. The figure shows the bulk momentum normalised to the energy band $\langle p \rangle/mc \gamma$ (top), and the beaming angle of the particle distribution normalised to $1/\gamma$ (bottom). Each point represents a calculation of the given quantity for all particles within the energy bin in a square of length $8\lambdap$.\label{fig:islandradius}}
 \end{figure}

  \begin{figure}[h]
\begin{center}
\includegraphics[width = 0.45\textwidth]{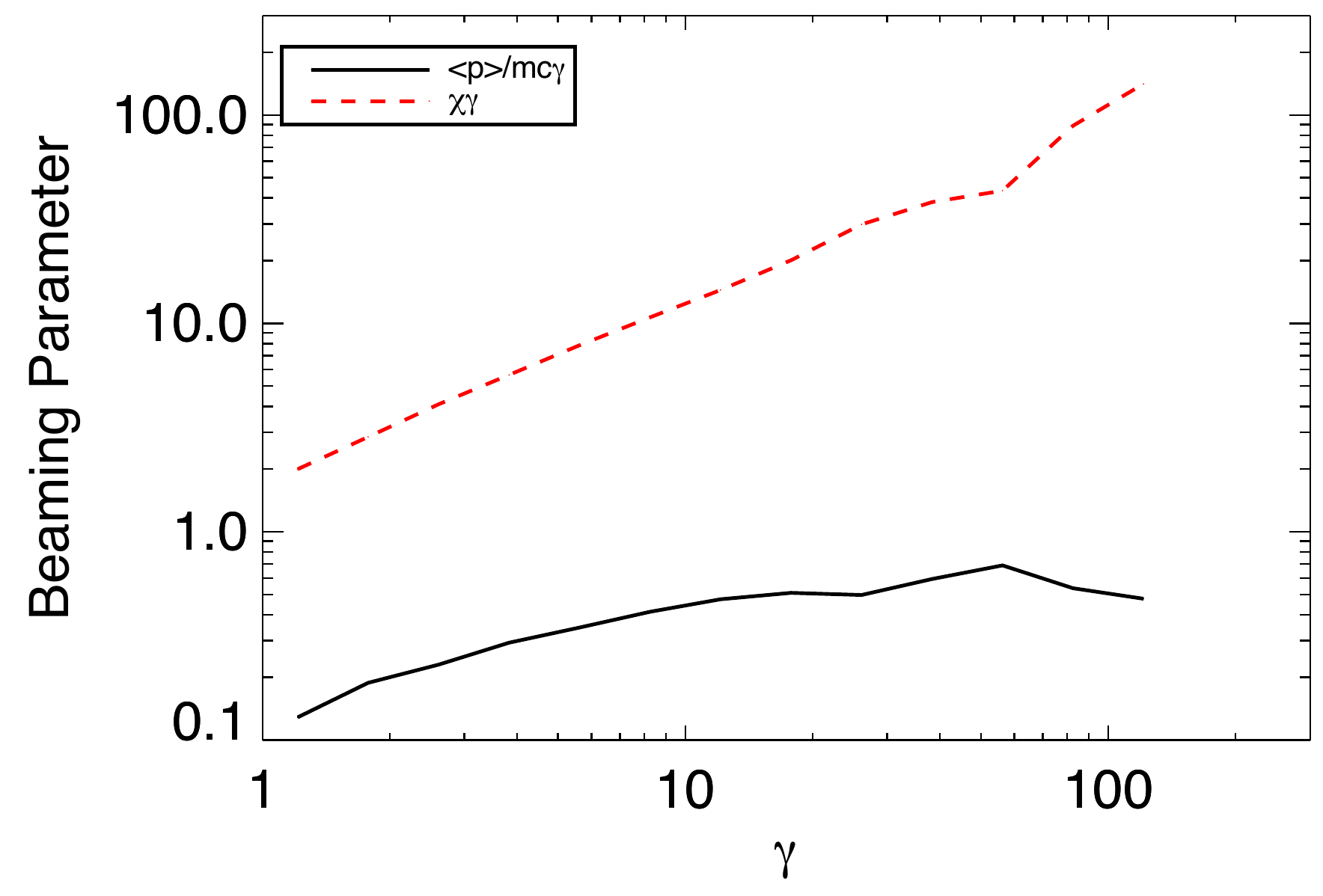}
\end{center}
\caption{ The variation with $\gamma $ of the average value of two beaming parameters in the magnetic island in Region C in simulation {\tt S40}.  The average is calculated from the values found in each square interval  weighted to the number of particles  in that spatial interval and energy bin. The figure shows the bulk momentum normalised to the energy band $\langle p \rangle/mc \gamma$, and the beaming angle of the particle distribution normalised to $1/\gamma$. \label{fig:islandgamma}}
 \end{figure}

We now investigate beaming in magnetic islands, focusing on Region C at $160<x/\lambdap<240$, $540<y/\lambdap<600$; we include particles that began in the current sheet, because they have a significant contribution to the total particle density at the island centers. Unlike the case in the X-points, the direction of the average particle velocity does not vary in a continuous fashion throughout the region. Flows in the $x-y$ plane tend to be towards the center of the island, but counterflows are often present as well. Similarly, the $z$ component of the average velocity tends to be oriented in the direction of particle acceleration in the X-points (the $+z$ direction for the positrons studied here), but it is reversed in some locations in the island. In general, particles are oriented relatively close to the $+z$ direction in the outlying areas of the islands that are not in the path of outflows from the X-points. 

While the direction of beaming of particles in the magnetic island  varies in irregular fashion, the magnitude of beaming varies with the radius from from the center of the island in a manner consistent with the increase in the magnitude of the overall bulk flows with radius (Figure \ref{fig:overallevol}, middle column, fourth row). The top panel of Figure \ref{fig:islandradius} shows that the normalized flow velocity increases with distance from the island center, but it is generally significantly smaller than the maximum values found in the X-point. The bottom panel shows that the normalised beaming angle decreases with radius, but it is still generally quite large at all radii. The outer regions in which this stronger beaming is present also contain higher typical particle energy (Figure \ref{fig:partspec}, bottom panel), so this result is consistent with earlier simulations that indicate that  particles that spend a long time in the island are thermalized and isotropized \citep{liu_11}, while particles that spend more time being accelerated in the X-point and enter the island later do not have time to undergo this process.

Figure  \ref{fig:islandgamma} shows the variation of the two parameters averaged over the whole island (Region C) with $\gamma$. It indicates that the normalised beaming angle $\chi \gamma$ increases quickly with $\gamma$, so that high-energy particles are not more strongly beamed than low-energy particles.  The normalised bulk momentum increases slowly with $\gamma$, but does not go above 0.8; smaller beaming angles correspond to a normalised momentum very close to 1, as shown in Figure \ref{fig:velmapping}.  The change in behavior of both parameters above $\gamma\sim 60$ is a result of the fact that the highest-energy particles in the islands are those that began in the current sheet. Such particles could began their acceleration earlier than those that began in the background plasma, but still have time to be isotropized in the islands.  Alternatively, they might have been reaccelerated during an island merger \citep{oka_10} or as result of island contraction \citep{drake_06}. 

       Overall, we do not find strong beaming of the particles anywhere in the magnetic islands except for the highest-energy particles in the regions where the outermost part of the island meets the reconnection outflows. Instead, we find a very moderate average beaming in the $+z$ direction, the  original direction of acceleration in the X-point. However, it is uncertain if this will mean there is no significant beaming of the resulting synchrotron radiation. Unlike the case of the X-points, where the majority of radiation is likely to be emitted where the magnetic field is $\sim B_0$ and the field varies slowly with $y$, the field in the island varies quickly with location. At $|\mathbf{d}-\mathbf{d}_i |=20 \lambdap$ the field can be as high as $\sim4B_0$, while the field is typically $\sim0.1 B_0$ at the center of the island. Thus, the radiation from the island may be highly beamed if specific locations on the outskirts of the island dominate the overall emission, which cannot be ruled out.  

\subsubsection{Schematic particle trajectory}

  \begin{figure}[h]
\begin{center}
\includegraphics[width = 0.45\textwidth]{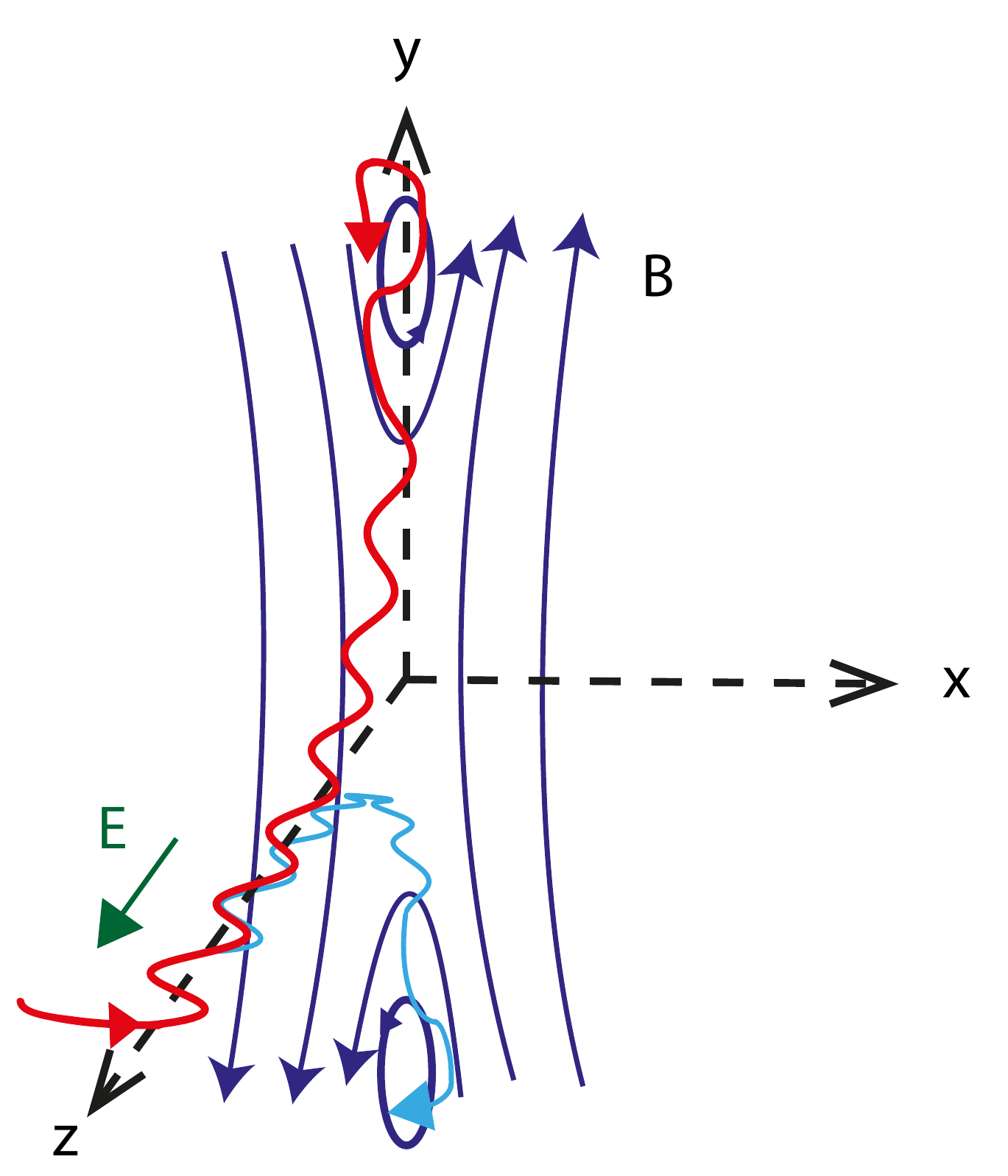}
\end{center}
\caption{  A schematic figure showing the trajectory of an electron (red) moving through an X-point and into an island in the $+y$ direction.  The characteristics of the trajectory correspond to the properties of the distribution of particles and the resulting beaming at each location. We also show a possible trajectory (light blue) in which the electron instead moves in the $-y$ direction to exit the X-point. The magnetic field structure is shown in dark blue. The number of oscillations is exaggerated for clarity; a typical electron only undergoes $1-2$ oscillations during its residence in the X-point.
\label{fig:reconschema}}
 \end{figure}
To clarify the conclusions of this section, we sketch a schematic particle trajectory and explain how it is consistent with the above conclusions. Figure \ref{fig:reconschema} shows the trajectory of an electron that enters the current sheet at large $z$. It oscillates across the current sheet in the $\pm x$ direction in a Speiser orbit  \citep{speiser_65} as it is accelerated by the electric field in the $-z$ direction. Eventually, it moves in the $+y$ direction towards a magnetic island and is accelerated in that direction as it moves. Thus, its velocity moves from the $-z$ direction towards the $+y$ direction as it moves towards the magnetic island in accordance with the top panel of Figure \ref{fig:velmapping}, except that the initial direction of acceleration is changed from $+z$ to $-z$ because the charge of the electron is negative. Once the electron reaches the island, it is moving mostly in the $+y$ direction, indicating that particles entering the island from the X-point are highly beamed. Afterwards, the electron is deflected into a randomised trajectory, indicating that the overall particle distribution in the magnetic island is not highly beamed, as shown in Figure  \ref{fig:islandgamma}. It remains mostly on the outskirts of the island, consistent with the finding in Figure \ref{fig:overallevol} that accelerated particles are mostly found in current sheets and the outskirts of magnetic islands.

\subsection{Beaming of particles and radiation}  \label{sec:beaming}

This section discusses the beaming of particles and radiation at various energies. Some differences exist between the beaming results in simulation {\tt S4} with $\sigma_0=4$ and those in the other two simulations due to the presence of the oscillations in the energy spectrum in the X-point at higher $\sigma_0$ discussed in \ref{sec:partaccel}. However, the oscillations do not have strong qualitative effects, so we focus on the results in simulation {\tt S40} with $\sigma_0=40$.

This section is organized as follows. We first show in Section \ref{sec:regimes} that the beaming in fast and slow cooling regimes can be probed by calculating the beaming of particles and the resulting radiation within an X-point and an island, respectively.  Then, in Sections \ref{sec:xpointmaps} and \ref{sec:islandmaps} we discuss the qualitative features of beaming maps in the X-point and the island. Section \ref{sec:quantbeam} compares quantitative measures of the beaming of particles and radiation for both regions with previous work. Finally, Section \ref{sec:schematicbeam} schematically summarizes the characteristics of radiation produced in the fast and slow cooling regimes.

\subsubsection{Fast and slow cooling regimes}\label{sec:regimes}
In our simulations, we do not directly determine the effect radiative cooling will have on the particles in a realistic reconnection configuration.  However, because particles can be accelerated without bound in the X-point during reconnection due to their focusing toward the center of the current sheet by electric acceleration \citep{uzdensky_11}, particles are unlikely to lose the bulk of their energy until they reach the edge of the X-point and enter the island. Therefore, the beaming characteristics of the radiation emitted by a particle will be determined by whether it cools before its momentum changes significantly in the magnetic field near the entrance to the island (fast cooling) or afterwards (slow cooling). Note that this cooling time is fast or slow compared not to the dynamical time for evolution of the overall reconnection structure but instead to the time for an individual particle to change its properties by motion within that structure.  If the highest-energy particles accelerated in the X-point are in the fast cooling regime, the beaming characteristics of particles and radiation at high energy are the same as those for the instantaneous particle and radiation distributions in Region B (an X-point). Otherwise all particles will radiate in the slow cooling regime, and the particles and radiation will be beamed similarly to the instantaneous distributions of the particles and radiation in Region C (a magnetic island). 

We now calculate the limiting Lorentz factor and emission energy at which the transition from slow to fast cooling occurs.  A particle with Lorentz factor $\gamma$ accelerated in an X-point with electric field $E$ gains energy at a rate
\begin{equation}
m c^2\left(\frac{d\gamma}{dt}\right)_{\rm accel}=q E c,\end{equation}
while the same particle radiating in a magnetic field $B$ loses energy at a rate
\begin{equation} m c^2\left(\frac{d\gamma}{dt}\right)_{\rm rad}=-\frac{2q^4 B^2 \gamma^2}{3 m^2 c^3}. \label{eq:coolingfirst}\end{equation}

The critical burnoff Lorentz factor at which the acceleration and cooling are equal is
\begin{equation}\gamma_{\rm bo}=\sqrt{\frac{3 m^2 c^4 E} {2 q^3 B^2}}.\end{equation}

We can rewrite Equation (\ref{eq:coolingfirst}) in terms of $\gamma_{\rm bo}$ as

\begin{equation}\left(\frac{d\gamma}{dt}\right)_{\rm rad}=-\frac{q E }{m c} \left(\frac{\gamma}{\gamma_{\rm bo}}\right)^2. \end{equation}
The cooling time can then be expressed in terms of the relativistic cyclotron frequency $\omega_{\rm g}=qB/(m c \gamma)$ as

\begin{equation}\omega_{\rm g}\ t_{\rm cool}= \frac{B}{E} \left(\frac{\gamma_{\rm bo}}{\gamma}\right)^2.\end{equation}

Because the time required for a particle's momentum to be modified significantly is just the particle's gyration time $1/\omega_{\rm g}$, a particle will be in the fast cooling regime if

\begin{equation}
\gamma>\sqrt{\frac{B} {E}} \gamma_{\rm bo}.
\end{equation}

The corresponding fast cooling limit for the characteristic photon energy may be found by combining this equation with the synchrotron frequency  Equation (\ref{eq:critomega}), setting $B_{\rm eff}=B$, and using the relation $\epsilon=h\omega$ for photons to obtain

\begin{equation}
\epsilon>\frac{9 m c^3 h} {4 q^2}\sim100\ {\rm MeV},
\end{equation}
where the latter estimate applies for physical electrons. 

This indicates that all emission below this energy will come from magnetic islands in the slow cooling regime.  As we see in the following sections, this has major implications for the beaming of this radiation.

 \begin{figure*}
\begin{center}
\includegraphics[width = \textwidth]{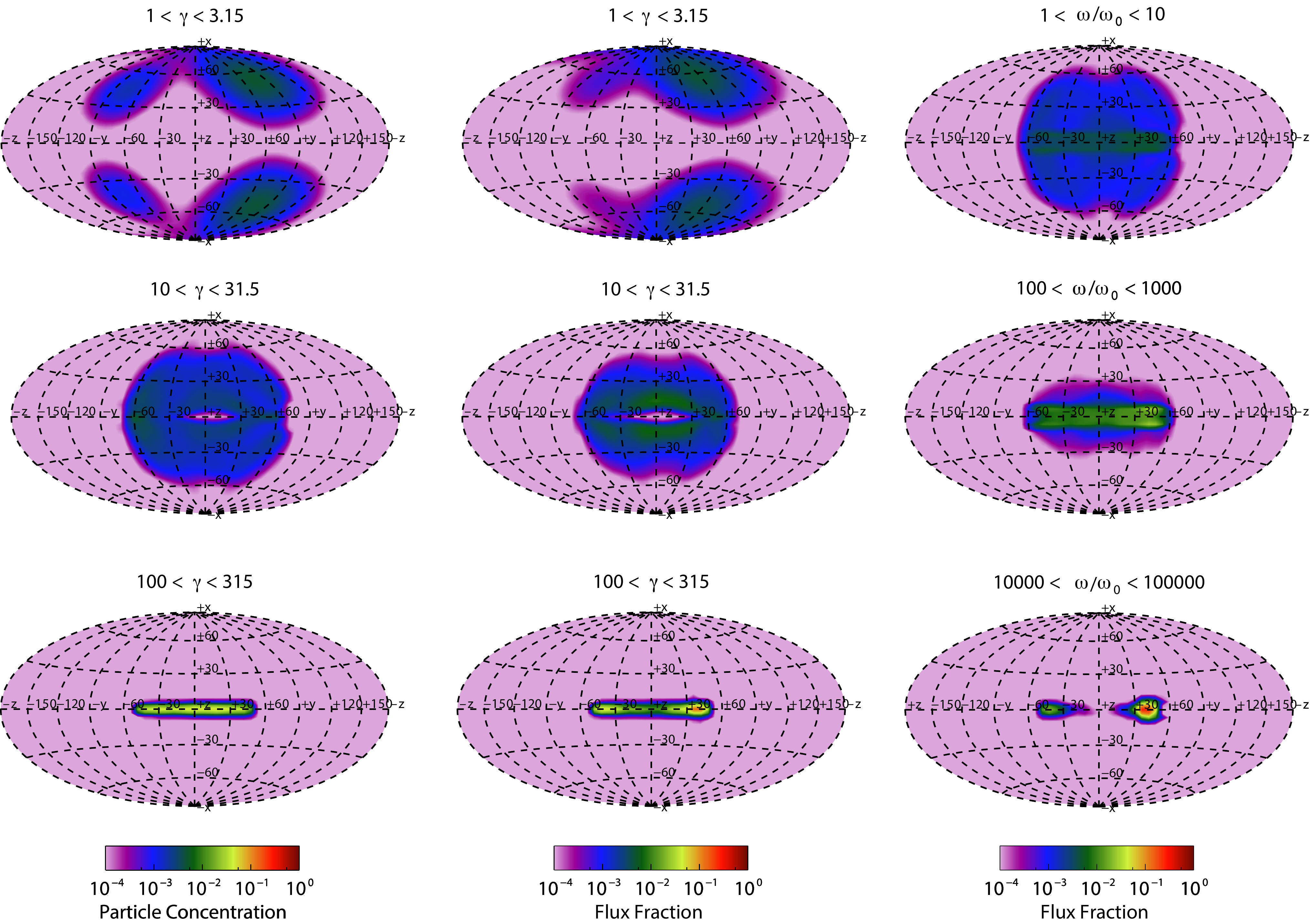}
\end{center}
\caption{ The angular distributions of the positrons (left column) in three ranges of energy, the total integrated synchrotron flux $\omega F_{\omega}$ from those positrons (middle column)  and the synchrotron flux $\omega F_{\omega}$ integrated over three logarithmic bands of frequency (right column) in Region B corresponding to an X-point at $184<x/\lambdap<216$, $320<y/\lambdap<520$ in simulation {\tt S40} in an Aitoff projection. Labels on each plot indicate the longitude $\phi$ or latitude $\theta$, or indicate that a point corresponds to a cartesian direction $(\pm x,\pm y, \pm z)$. Each plot is normalised to the total number of positrons (left column) or the total integrated synchrotron flux (middle and right columns). The distributions in this region correspond to a fast cooling regime. \label{fig:xpointmaps}}
\end{figure*}

\subsubsection{Beaming maps in the X-point}\label{sec:xpointmaps}

In both regions, we calculate the angular distribution of particles and of the synchrotron radiation in three energy bands. The angular distribution is given as a function of the longitude $\phi=\tan^{-1} (y/z)$ (the angle within the current sheet) and latitude $\theta=\sin^{-1} x$ (the angle perpendicular to the current sheet), where $\mathbf{r}=(x,y,z)$ is the unit vector corresponding to a given direction. The frequency $\omega$ of radiation is normalized to $\omega_0=0.29 \omega_c$, the fiducial peak synchrotron frequency of a particle with $\gamma=1$ moving perpendicular to the background magnetic field $B_0$. Thus, the radiation spectrum from a particle with Lorentz factor $\gamma$ is expected to peak at $\omega/\omega_0=\gamma^2$ and the particles in each interval of Lorentz factors are expected to emit in the corresponding radiation energy band. However,  particles at higher energy in each range often contribute more synchrotron radiation in each band than those at low energy do, and many particles emit at significantly higher or lower magnetic field than $B_0$, so that the energy ranges do not precisely correspond.  Therefore, we also calculate the angular distribution of the total synchrotron flux as a function of direction from particles in each interval.

Figure \ref{fig:xpointmaps} shows Aitoff projections of the angular distribution of the positrons in three energy bands, the total synchrotron flux from those particles, and the synchrotron flux from all particles $\omega F(\omega)$ integrated over three logarithmic energy bands in Region B, corresponding to an X-point. This radiation corresponds to a fast cooling regime. In the X-point, the particle energy distribution $\gamma dN/d \gamma $ shown in Figure \ref{fig:partspec} is nearly flat above $\gamma\sim 20$, so that the synchrotron energy per logarithmic interval in the two lower frequency ranges is dominated by the emission from particles at significantly higher $\gamma$. Therefore, the particle and radiation angular distributions are significantly different for the two lowest energy bins.

\begin{figure*}
\begin{center}
\includegraphics[width = \textwidth]{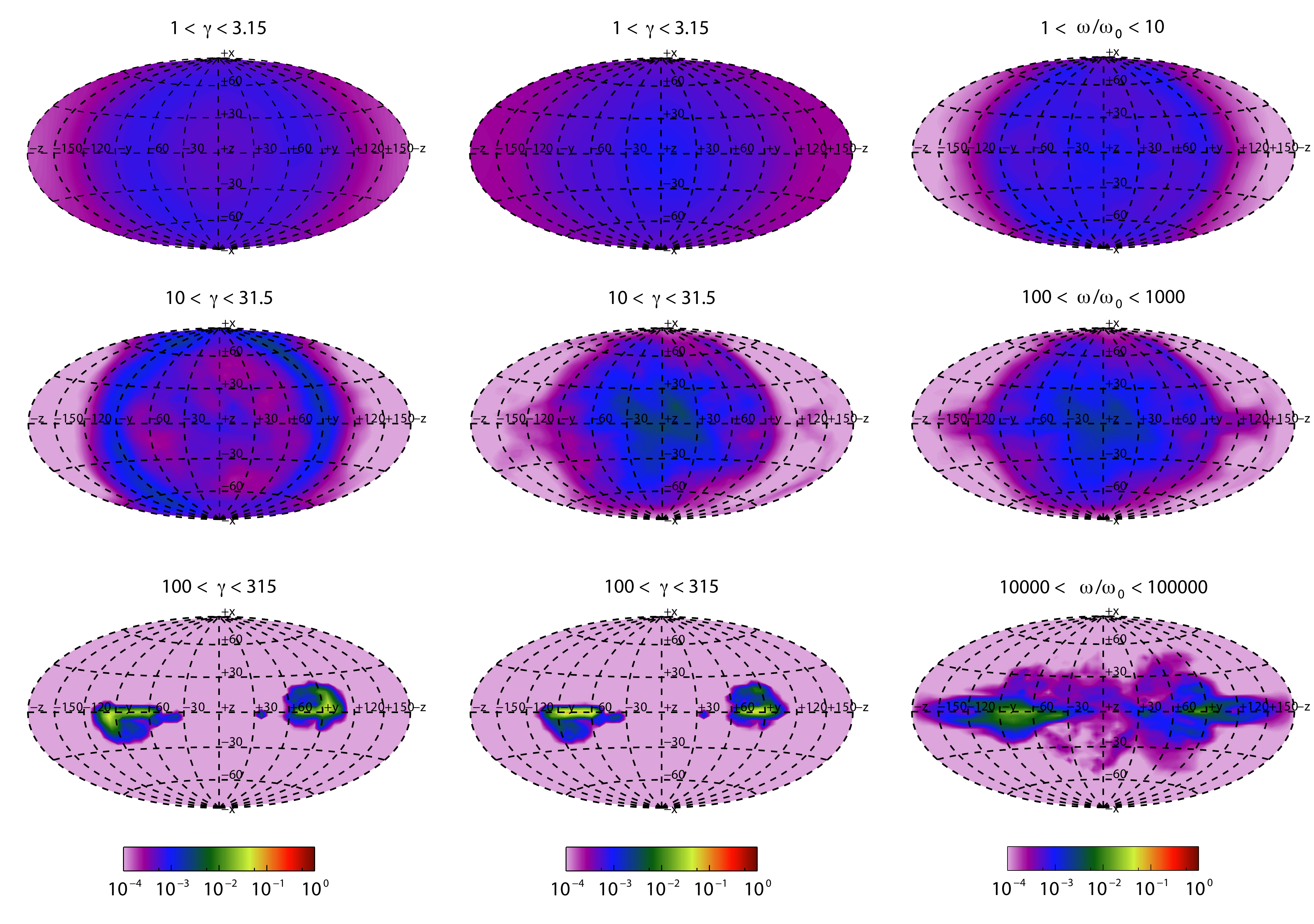}
\end{center}
\caption{ The angular distributions of the positrons (left column) in three ranges of energy, the total integrated synchrotron flux $\omega F_{\omega}$ from those positrons (middle column)  and the synchrotron flux $\omega F_{\omega}$ integrated over three bands of frequency (right column) in Region C (a magnetic island) in simulation {\tt S40} in an Aitoff projection.   Labels on each plot indicate the longitude $\phi$ or latitude $\theta$, or indicate that a point corresponds to a cartesian direction $(\pm x,\pm y, \pm z)$. Each plot is normalised to the total number of positrons (left column) or the total integrated synchrotron flux (middle and right columns). The distributions in this region correspond to a slow cooling regime. \label{fig:islandmap}}
\end{figure*}

The particle distribution in the  lowest energy band (top row, left column), corresponding to  $1<\gamma<3.15$  (and the synchrotron radiation from particles in that range, middle column) is highly anisotropic, but the percentage of particles that are at such a small $\gamma$ in the X-point is extremely low (Figure \ref{fig:partspec}), and they do not contribute significantly to the synchrotron radiation at any frequency. Comparison of the second and third rows of Figure \ref{fig:xpointmaps}  indicates that the particles (left column) and the resulting radiation (middle column) become more anisotropic as the energy increases, with all of the distributions becoming focused towards the equator at $\theta=0$. The particles are concentrated in two populations. The first is located at $\theta\sim 0$ and located approximately at the value of $\phi$  predicted from  $\tan^{-1} (p_y/p_z)$ calculated at the edge of the X-point. For the range $10<\gamma<31.5$  ($100<\gamma<315$)  these particles are located at  $\phi \sim \pm60 \degrees\ (\pm 50 \degrees)$, which is close to the value of $\tan^{-1} (p_y/p_z) \approx 58 \degrees (45 \degrees)$ for $\gamma \sim 26 \ (120)$ found in Figure \ref{fig:velmapping} for particles in the outflow regions at the edge of the X-point.  The second  large group of particles has a significant velocity in the $\pm x$ direction but is mostly moving in the $+z$ direction in the $y-z$ plane, with its motion characteristic of Speiser orbits in the center of an X-point.  For $100<\gamma<315$, the particles moving in the $+z$ direction are also in Speiser orbits, but they have a much smaller momentum in the $\pm x$ direction because they have spent more time being accelerated \citep{uzdensky_11}.  These particles are expected to produce little radiation because they spend much of their time in the current sheet where the magnetic field is low, and indeed the expected radiation for this population (middle column, bottom row) is small.

The synchrotron flux distributions in the three frequency bands are significantly different than those of the particles as a result of the sharply peaked energy distribution of the particles.  The angular distribution in the frequency band $1<\omega/\omega_0<10$ is clearly similar to that of the particles in the next higher energy band  $10<\gamma<31.5$. For $ 100<\omega/\omega_0<1000$ two components are present. One has relatively low amplitude and is similar to the distribution of the corresponding particles at $10<\gamma<31.5$. The second is concentrated only along the equator with a smaller range of $\phi$ and has a significantly higher amplitude.  This radiation is likely produced by higher-energy particles moving in Speiser orbits, which peak in this frequency band because they radiate in significantly lower magnetic field in the X-point. Finally, for $ 10000<\omega/\omega_0<100000$ the radiation is extremely beamed  on the equator at $\phi \sim \pm 45 \degrees$. This beaming is even stronger than that expected from the distribution of high-energy particles, but can be explained by considering that slightly lower-energy particles will only contribute to this frequency band if they are at the edge of the X-point where the magnetic field is close to $B_0$ and the beaming is in this direction. Overall, in the fast cooling case in which most emission comes from the edge of the X-point, we expect the observed radiation to be highly beamed.

\begin{figure*}[t]
\begin{center}
\includegraphics[width = 0.9\textwidth]{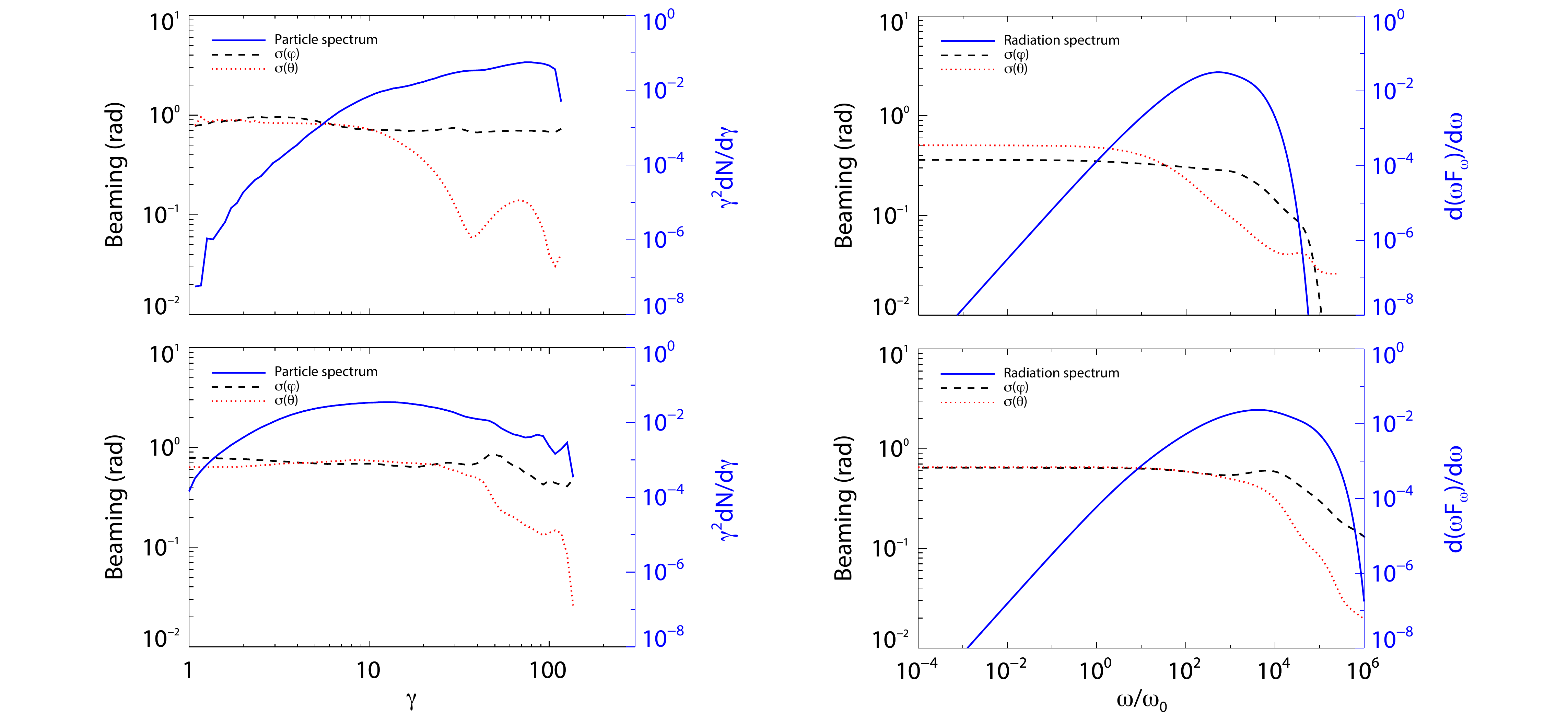}
\end{center}
\caption{ The total positron energy spectrum (left) and synchrotron flux spectrum $\omega F_{\omega}$ (right)  as well as the standard deviations of the directional distribution in the $\phi$ and $\theta$ directions as a function of $\gamma$ (left) and $\omega/\omega_0$ (right) . These are calculated for Region B in an X-point, which corresponds to the fast cooling regime (top row),  and Region C in an island, which corresponds to the slow cooling regime with most cooling occurring after the end of reconnection (bottom row). Note that the calculations in the islands include the particles that began in the current sheet, which is why the distribution there continues to slightly higher energies than those found in the X-point. \label{fig:spreads}}
 \end{figure*}
 
\subsubsection{Beaming maps in the island}\label{sec:islandmaps}
Figure \ref{fig:islandmap} shows Aitoff projections of the angular distribution of the positrons in three energy bands, the total synchrotron flux from those particles, and the synchrotron flux from all particles $\omega F(\omega)$ integrated over three logarithmic energy bands in Region C, corresponding to a magnetic island.  We include particles that began the simulation in the current sheet in our analysis, because they can have a significant portion of the total energy at the center of the island.  This radiation corresponds to a slow cooling regime in which radiation is emitted well after particles exit the X-point and enter the island. The distributions are significantly simpler than those found in the X-point.  

At low energies, all three distributions are nearly isotropic, although there is a slight concentration of synchrotron flux in the $+z$ direction. For the middle energy band, the particle distribution for $10<\gamma<31.5$ is concentrated so that most particles have no significant momentum in the $\pm z$ directions. It is uncertain why this is the case, but the concentration is in any case  fairly small and a significant number of particles are oriented in the $+z$ direction. The  total synchrotron radiation from these particles and the synchrotron radiation in the corresponding energy band $ 100<\omega/\omega_0<1000$ are more isotropic, with a slight concentration in the $+z$ direction that may be explained by the fact that the locations where most particles are oriented in the $+z$ direction are generally in the outlying areas of the island, where the magnetic field is strong. The particles in the highest energy range $100<\gamma<315$ are highly anisotropic and oriented in the $\pm y$ direction, and the radiation from these particles has an identical distribution. These are clearly the highest-energy particles from the population entering the island from the outflow regions of the X-points, which are shown in Figure \ref{fig:partspec} to have the highest average energy in the islands. Note, however, that these particles are not as concentrated in the $\theta$ direction as the highest-energy particles in the X-point. The synchrotron radiation in the corresponding frequency band $10000<\omega/\omega_0<100000$ is considerably more spread out in both $\phi$ and $\theta$,  probably as a result of the radiation of lower-energy particles in locations with large magnetic field $>B_0$ that contribute significantly to the observed distribution.  

Overall, we expect the emission from islands to be relatively unbeamed except at the highest energies. This high-energy radiation comes from particles located where outflows from X-points meet the magnetic islands. However, in the physical slow cooling case this high-energy particle population will probably not radiate until its momentum has been significantly altered by the field of the magnetic island. If this is true, we expect no significant beaming at any frequency for this slow-cooling case. Because AGN and GRBs emit at energies significantly below $160$ MeV in the slow cooling regime, models of these systems that require strong beaming may need to be reconsidered. However, this conclusion holds only if acceleration is a single-step process. If significant reacceleration that produces beaming occurs in magnetic islands, radiation may be strongly beamed below this limit. Because particles in the slow cooling regime defined in this paper can still cool quickly compared to the dynamical time of the overall reconnecting region, there is a regime in which reaccelerated particles radiate efficiently below 160 MeV.  However, it is not clear that acceleration mechanisms that work in magnetic islands, such as island contraction \citep{drake_06} and island coalescence \citep{oka_10} are likely to produce strong beaming. 

\subsubsection{Quantitative beaming calculations and comparison to other work}\label{sec:quantbeam}

We now make more quantitative comparisons of the beaming of particles and radiation in both regions. Because the angular distributions of the particles in a given energy band and the total emission from those particles (left and middle rows of the previous figures) are very similar, we only compare the particle beaming as a function of $\gamma$ to the synchrotron distribution from all particles in a region as a function of $\omega/\omega_0$.  Because all of the distributions appear to have much larger spread in $\phi$ than in $\theta$ at high energies, we calculate two standard deviations $\chi(\phi)$ and $\chi(\theta)$ of the angular distributions about their means $\phi_m$ and $\theta_m$ instead of using a single parameter; we discuss the method of calculation in an Appendix. To correct for the presence of double peaks in the angular distributions for opposite signs of $\phi$, we restrict our calculation to the region $\phi<0$. 

Figure \ref{fig:spreads} shows the variation of the standard deviations $\chi(\phi)$ and $\chi(\theta)$ of the angular distributions about their means $\phi_m$ and $\theta_m$  as a function of energy and the spectral energy distribution for both positrons and their synchrotron radiation in both regions. The left column of the figure shows that the value of $\chi(\phi)$ for the particles is insensitive to $\gamma$ in both regions (although there is a slight decrease in this parameter in the island at large $\gamma$),  and that $\chi(\phi)$ for the radiation is also insensitive to $\omega/\omega_0$ below the peak of the synchrotron spectrum, which is typically at $\omega/\omega_0 \sim 10000-100000$. Above this peak, $\chi(\phi)$ declines with increased frequency $\omega/\omega_0$ in both regions, but only becomes comparable to the small values of $\sim .05$ rad found for $\chi(\theta)$ in the case of the X-point. In the X-point (Region B), the reason for this decline is probably that particles at high energies that are moving in the $+z$ direction can no longer contribute significantly above the peak because they are in regions of low magnetic field. In the island (Region C), the reason for this decline and the slight decline in the particle beaming is probably that the particle population that is entering the island from the adjacent X-points becomes more prominent, and there is less contribution from particles oriented in the $+z$ direction. 

  \begin{figure*}[t]
\begin{center}
\includegraphics[width = 0.9\textwidth]{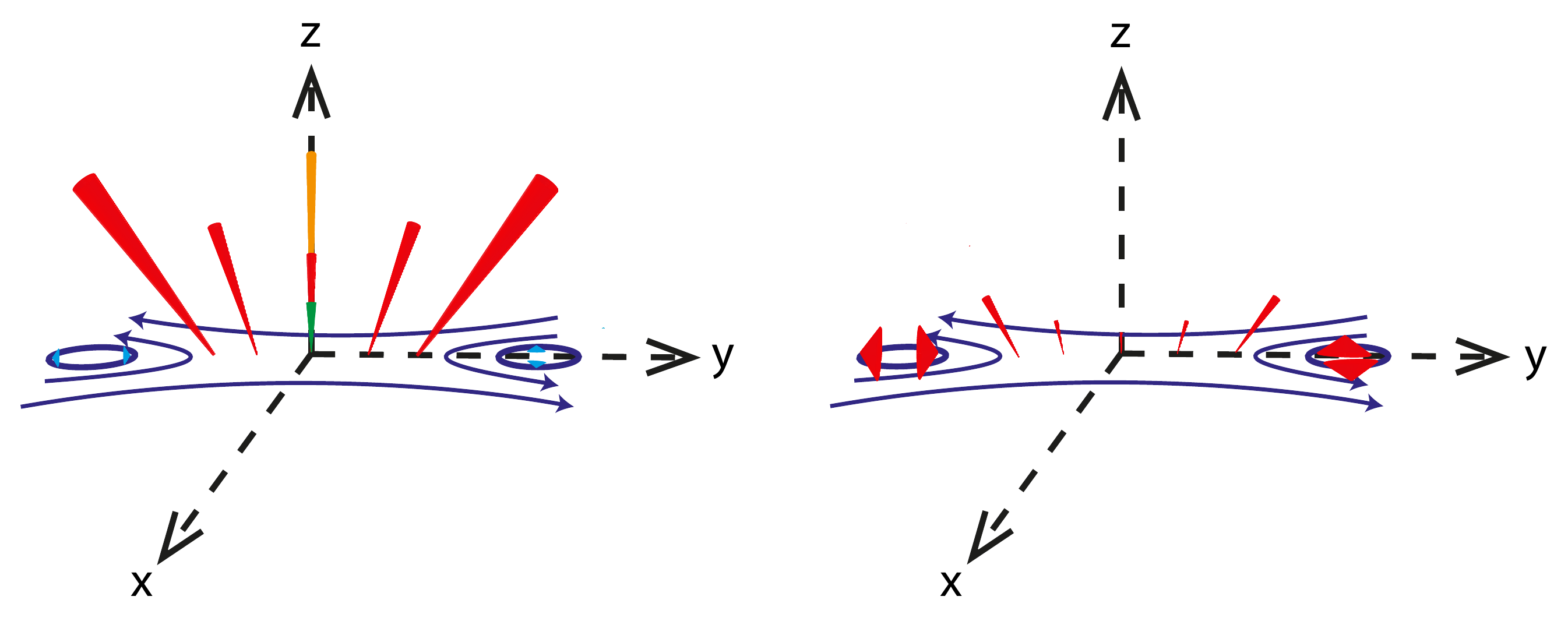}
\end{center}
\caption{  A schematic figure showing the beamed radiation resulting from positrons with $\gamma=50$ at various locations in the current sheet in the fast cooling (left) and slow cooling (right) cases. The magnetic field structure is shown in dark blue. Each cone indicates the solid angle within within which most of the radiation is beamed at a given location. The volume of the cone approximately indicates the amount of radiation emitted by each particle at that location, but the scales are different in the fast and slow cooling cases.  To illustrate how these properties vary with $\gamma$, we also show cones in green and orange at a single location for particles with $\gamma=25$ and $\gamma=100$, respectively, in the fast cooling case. The cyan cones in the islands in the fast cooling case represent emission from particles with $\gamma$ of order unity. \label{fig:beamingschema}}
 \end{figure*}

$\chi(\theta)$ depends much more strongly on energy for both the particles and the radiation. In the X-point, $\chi(\theta)$ declines with $\gamma$ for $\gamma>10$, but an oscillatory signature is present in which less spread is present at locations corresponding to the troughs present in the particle energy distribution for Region A in Figure \ref{fig:partspec}, and more spread is present at other locations. The oscillatory signature is also present in the radiation from the X-point, producing a slight rise in spread from $\omega=10000-100000$, but it is much less significant because the particle energy distribution in the X-point is strongly peaked at $\gamma\sim 100$ and most of the radiation comes from there.  In the absence of the oscillation, we expect the dips in spread of the particles to disappear, but it is difficult to calculate the expected trends. For particles moving on Speiser orbits with amplitude larger than the current sheet width, \citet{uzdensky_11} predicts that the spread should decrease as $\gamma^{-2/3}$, which is roughly consistent with the decline from $\chi(\theta)=0.8$ at $\gamma=10$ to $\chi(\theta)=0.15$ at the peak in spread at $\gamma\sim 75$. In the case of the radiation, the value of $\chi(\theta)$ decreases approximately as $\omega^{-1/3}$ for  $10<\omega/\omega_0 <10000$, which is again roughly consistent with the expected results from Speiser orbits because $\omega \propto \gamma^2$. A decline in $\chi(\theta)$ with $\gamma$ also occurs in the island, but only for $\gamma>60$ above the peak of the particle energy distribution, and it only decreases to $\sim 0.15 $ rad at $\gamma=100$; the steep decline at $\gamma>100$ may be a result of small number statistics. The corresponding decline in $\chi(\theta)$ with $\omega$, which is approximately proportional to $1/\omega$, begins above the peak of the radiation distribution at $\omega/\omega_0=50000$ and continues to high energies, but it ls much less steep than the $\gamma>100$ decline in the particle energy distribution. Again, sharp features in the particle beaming are smoothed out in the radiation beaming.

We now compare our results for the beaming of radiation with other work. Because we are interested in high energy radiation that contains significant portion of the total energy, we focus on the values of $\chi(\theta)$ and $\chi(\phi)$ one decade above the peak of the distribution in each region. which is approximately at $\omega/\omega_0=5000$ for Region B (the X-point), and $\omega/\omega_0=50000$ for Region C (the island). The beaming in the simulations of \citet{cerutti_12b} without radiative feedback, and \citet{cerutti_13a, cerutti_14}, which include radiative feedback, indicates that high energy radiation within one decade in frequency above the peak in $\omega F_{\omega}$ is beamed within a solid angle of $\Omega/4\pi\sim 0.02-0.04$, The solid angle within which the particles with $\phi<0$ are beamed in our simulations may be estimated as $\pi \chi(\theta)\chi(\phi)$, so the beaming solid angle for all $\phi$  in our simulation {\tt S40} is $\Omega /4 \pi \sim  \chi(\theta)\chi(\phi)/2$, where the standard deviations are expressed in radians in the equation. In Region B (the X-point), at  $\omega/\omega_0=5000$,  $\chi(\theta)= 0.05$ rad, $\chi(\phi)=0.18$ rad, and $\Omega/4\pi \sim0.005$. In Region C (the island) at  $\omega/\omega_0=50000$, $\chi(\theta)= 0.12$ rad, $\chi(\phi)=0.35$ rad, and $\Omega /4 \pi \sim 0.02$. Thus, our results indicate that strong beaming is present at high energies in both cooling regimes, and our results are consistent with previous work but indicate slightly stronger beaming. However, in the slow cooling case the narrowing of the spread at high energies due to particles entering the island from the X-points may not occur because radiation is produced well after these particles leave the X-points and are deflected by the island's magnetic field. In this case, the spread at high energies will remain similar to that at low energies with  $\chi(\theta)=\chi(\phi)=0.7$ rad, and $\Omega/4\pi\sim0.25$. Thus, it possible that high-energy radiation produced in relativistic magnetic reconnection is strongly beamed only in the fast-cooling case.

\subsubsection{Schematic beaming of radiation} \label{sec:schematicbeam}
To clarify the conclusions of this section, we schematically show in Figure \ref{fig:beamingschema} the magnitude and opening angle of the radiation coming from particles that are accelerated to $\gamma \sim 50$ in the fast cooling and slow cooling cases, although. In both cases, the opening angle of the radiation distribution from these particles in the X-point is $\sim 5/\gamma=0.1$ rad. The figure for the fast cooling case also illustrates that the amount of radiation increases and the opening angle decreases as $\gamma$ is increased. In the X-point, the direction of beaming varies from the direction of acceleration ($+z$) towards the direction of the outflow towards the magnetic island ($\pm y$) as found in our analysis of particle beaming in Figure \ref{fig:velmapping}.  The amount of radiation emitted by each particle increases with distance from the center of the X-point due to the variation of the $x$ component of the magnetic field.  In contrast, in the magnetic islands, the opening angle of the radiation distribution is very wide, approximately 1 rad as seen in Figure \ref{fig:islandgamma}, and the direction of beaming varies greatly with location. We show four representative beaming directions at various locations in the islands.

In the fast cooling case (left), each particle loses most of its energy at the edge of the X-point in a very short time, and emits little energy thereafter. The blue cones in the islands correspond to emission from particles with $\gamma$ of order unity, which is relatively unimportant.  This means the high-energy radiation in the fast cooling case is similar to the instantaneous emission from the X-point, which is dominated by particles at its edge. In the slow cooling case (right), particles lose very little energy before reaching the island. Because they spend most of their time in the island and the magnetic field is high there, the radiation emitted from the island dominates the observed radiation. This is again in agreement with our previous statements that the islands are responsible for most emission in the slow cooling case. Note that in the dynamical simulations of \citet{yuan_16}, the low radiative efficiency in the slow cooling case is likely because the emission of the particles during the dynamical phase is much less than that which will occur afterwards in the quiescent phase, consistent with this conclusion.
\section{Conclusions}\label{sec:conclusions}
We have carried out particle-in-cell simulations of relativistic magnetic reconnection at three values
of the  background magnetisation $\sigma_0$= 4, 40,  and 400, and calculated the particle and synchrotron radiation
beaming resulting from the reconnection process. Given the limits of our simulations not all these results are 
necessarily generic to systems of larger size that exist over longer durations or for systems with larger $\sigma$. Therefore we first summarize our 
findings in these specific simulations we carried out and then we identify the results that we expect to be generic and relevant for astrophysical systems such as GRBs and AGNs.

Our simulations are characterized by the following properties:
\begin{itemize}
\item  The reconnection rate and the characteristics of the
angular distributions of particles and radiation do not
change significantly with the background magnetisation
$\sigma_0$, despite the significantly thicker current sheet structure
and the much higher inflow velocities in low density
plasmas with high values of $\sigma_0$ that are produced
for our choice of initial conditions. This indicates
that most of the important aspects of relativistic
reconnection do not depend strongly on the background
magnetisation so long as it is larger than 1, and that
the choice of an initial equilibrium with high density
contrast at high magnetization does not change these
aspects of reconnection either. 

\item Particle acceleration in our simulations is efficient,
producing a hard power-law tail in the particle
energy spectrum of the form $dN/d\gamma \propto \gamma^{-\alpha}$ 
- with index $\alpha \approx 1.5$ that becomes harder as $\sigma_0$ increases. The
high-energy cutoff $\gamma_p$ of the power law distribution also
increases with $\sigma_0$ in accordance with energetic constraints.

\item Particles are accelerated into this
power-law distribution primarily in X-points, with no significant acceleration taking place within the islands. 
Once particles are accelerated they leave the X-point and move towards the islands. In our simulations, which are limited in size and time, the highest energy particles still reside at the end of the simulations in the X-points, while particles with $\gamma \approx \gamma_p/2$ are just entering the islands.

\item The outflow of accelerated plasma from the X-points has a complicated beaming pattern and cannot be regarded simply as a relativistic bulk flow - as commonly envisaged in idealized analytic models. In fact the bulk outflow, $\langle p \rangle /mc\gamma$, is mildly relativistic for all values of $\sigma_0$ we studied. Instead the beaming size is $\gamma$ dependent and beaming directions vary from one position to another. Particles with Lorentz factor $\gamma$ are typically beamed within $\sim 5/\gamma$. The direction of the beam rotates with location from the direction of the electric field in the center of the X-point to the direction of outflow towards the island at the edge of the island. The result of this rotation is that the integrated population of particles with a Lorentz factor $\gamma$ that are still in the X-points (i.e., haven't entered the islands yet) forms a fan narrowly confined to the plane of the current sheet (the y-z plane). Once entering the islands, the accelerated particles gradually lose their beaming, so the island population is mostly unbeamed. 

\item The resulting synchrotron radiation pattern follows suit. Radiation from the X-points is highly beamed. Since the magnetic field in the center of the X-point is weak radiation is significant only from particles that are leaving the X-points towards the island. As a result the narrow fan of particles produces a radiation pattern that is dark in the center and becomes brighter towards the edges of the fan. Because the fan width depends on $\gamma$ higher energy radiation is more narrowly beamed. In our simulation the total beaming at the  highest synchrotron frequencies of radiation coming from the X-points is $0.5\%$ of the sky. In magnetic islands, high-energy particles are unbeamed of so is the radiation they generate. An exception is the radiation from particles that are just entering the islands from the X-points and did not have time to isotropise. In our simulation, where the highest energy particles did not have time to isotropise in the islands, their radiation in the strong magnetic field of the islands produces a strongly beamed synchrotron component from the islands themselves. 

\end{itemize}

Based on these results we can draw some generic conclusions that are relevant for astrophysical settings. Most and possibly all particles are accelerated in the X-points and flow into the islands where they spend most of their life. Exceptions are particles that cool down radiatively within the X-points. The radiation from the X-points is beamed into a narrow fan while radiation from the islands is not strongly beamed. The combination of the beaming pattern from different regions and the fact that most particles spend most of their lifetime in islands implies that synchrotron radiation from reconnection sites is generally rather isotropic. Highly beamed radiation may be generated only by particles that cool down rapidly before reaching the islands.  As we have shown these particles emit around or above the burn-off limit, which is $\sim 100$ MeV in the rest frame of the reconnection site.  Therefore, models of GRBs and AGN that require strong beaming at lower energies may
need to be reevaluated. 
\acknowledgements 
We thank L. Sironi for useful discussions of the overall evolution of these simulations in comparison to his,  M. Hoshino for discussions of the possible effects of 3D physics on our simulations without guide field and suggestions regarding the physics of fast inflows at high magnetization, and P. Binyamini and J. Granot for discussions of the applications of magnetic reconnection to the jet-within-a-jet model of GRBs. TP was supported by the I-CORE Center for Excellence in Research in Astrophysics, and a  CNSF-ISF grant. DK and EN were partially supported by an ERC starting grant (GRB/SN), ISF grant (1277/13) and an ISA grant.

\appendix
Here, we discuss our calculation of the directional standard deviation $\chi$ of the particles in the X-points and islands in Section \ref{sec:velmapping}, and the two-dimensional directional standard deviations $\chi(\theta)$ and $\chi(\phi)$ for both particles and radiation in Section \ref{sec:beaming}.

 To calculate $\chi$, we assume that the directions of the particle momenta are distributed in a von Mises-Fisher distribution \citep{fisher_53} of the form $f(\xi)\propto \exp(-\kappa \cos \xi)$,  where $\kappa$ is the concentration parameter that represents how clustered the directions are, and $\xi$ is the spherical angle relative to the mean direction. This distribution is the spherical equivalent of the  Gaussian distribution. Treating the direction of motion as a vector $\mathbf{r}=(x,y,z)$ on the unit sphere, the confidence interval within which the mean direction is $68\%$ likely to fall is given by
   \begin{equation}
\chi=\cos^{-1}\left(\frac{\ln((1-0.68) e^{2\kappa}+0.68)}{\kappa}-1 \right),
 \end{equation}
 
 where $\kappa$  may be approximated as \citep{banerjee_05}
 
    \begin{equation}
\kappa=\frac{R(3-R^2)}{1-R^2},
 \end{equation}
  
 and 
 
   \begin{equation}
 R=\frac{\sum_i \mathbf{r}_i}{N}.
 \end{equation}
 
 Here $R$ is the ratio of the summed vector to the number of vectors, which is equivalent to the normalised momentum calculated above but for directions only, and $i$ indicates a summation over all particles. We take this value of $\chi$ to be an accurate estimate of the spread of the particle direction about the mean.

Under the assumption that the angular distribution is a bivariate von Mises-Fisher distribution in the $\theta$ and $\phi$ directions, accurate confidence intervals for the uncertainty in the mean direction may be calculated merely by calculating the standard deviations in the corresponding cartesian coordinates $\sin(\phi-\phi_m)$ and $\sin(\theta-\theta_m)$ \citep{kent_82}

  \begin{equation}
 \chi(\alpha)=\sin^{-1} \sqrt{\frac{\int F \sin^2 (\alpha-\alpha_m)d \Omega}{\int F d\Omega} },        \alpha=\theta, \phi
 \end{equation}
 where $F$ is the number of particles or the amount of synchrotron flux emitted in a given direction within the energy band, and $d\Omega$ is the surface area element. We again take the confidence intervals to be accurate estimates of the spread in direction about the mean.

\end{document}